\documentclass[aps,prd,
preprint,
floatfix,nofootinbib,groupedaddress,showpacs]{revtex4}
\usepackage{graphicx}
\usepackage{amsmath}
\usepackage{latexsym}
\usepackage{here}
\usepackage{slashbox}
\usepackage{array}
\usepackage{dcolumn}% Align table columns on decimal point
\usepackage{bm}% bold math
\usepackage{epsfig}
\usepackage{multirow}

\usepackage[dvips]{color}
\definecolor{Black}{named}{Black}
\definecolor{Red}{named}{Red}
\definecolor{Blue}{named}{Blue}

\newcommand{\Lumint}{{\cal L}_{\rm int}}

\def\epem{\ifmmode e^+e^-\else $e^+e^-$\fi}
\def\to{\rightarrow}

\def\mpl{\ifmmode \overline M_{Pl}\else $\bar M_{Pl}$\fi}
\def\beq{\begin{equation}}
\def\be{\begin{equation}}
\def\beqn{\begin{eqnarray}}
\def\ee{\end{equation}}
\def\eeq{\end{equation}}
\def\eeqn{\end{eqnarray}}

\begin{document}

\title{Discriminating  $Z^\prime$ from anomalous trilinear gauge coupling signatures
in $e^+e^- \rightarrow W^+W^-$  at  ILC with polarized beams}

\author{
V. V. Andreev,$^{a,}$\footnote{quarks@gsu.by}\hspace{.4cm} 
G. Moortgat-Pick,$^{b,}$\footnote{gudrid.moortgat-pick@desy.de}\hspace{.4cm}
P. Osland,$^{c,d,}$\footnote{per.osland@ift.uib.no}\hspace{.4cm} 
A. A. Pankov$^{e,}$\footnote{pankov@ictp.it}\hspace{.4cm} 
N. Paver$^{f,}$\footnote{nello.paver@ts.infn.it}\hspace{.4cm} }
\vspace{.5cm} \affiliation{
 $^{\rm a}$The F. Scorina Gomel State University, 246019 Gomel, Belarus\\
$^{\rm b}$DESY FLC,
Notkestrasse 85, Hamburg 22607, Germany\\
$^{\rm c}$Department of Physics and Technology, University of Bergen, Postboks 7803, N-5020  Bergen, Norway\\
$^{\rm d}$CERN, CH-1211 Gen\`eve 23, Switzerland\\
 $^{\rm e}$The Abdus Salam ICTP Affiliated Centre, Technical University of Gomel, 246746 Gomel,
 Belarus\\
 $^{f}$University of Trieste and INFN-Trieste Section, 34100 Trieste, Italy
 }
%\date{\today}

\begin{abstract}
New heavy neutral gauge bosons $Z'$ are predicted by many models
of physics beyond the Standard Model. It is quite possible that
$Z'$s are heavy enough to lie beyond the discovery reach 
of the CERN Large Hadron Collider LHC, in which case only indirect
signatures of $Z^\prime$ exchanges may emerge at future colliders,
through deviations of the measured cross sections from the
Standard Model predictions. We discuss in this context the
foreseeable sensitivity to $Z'$s of $W^\pm$-pair production cross
sections at the $e^+e^-$ International Linear Collider (ILC),
especially as regards the potential of distinguishing  observable
effects of the $Z'$ from analogous ones due to competitor models
with anomalous trilinear gauge couplings (AGC) that can lead to the same
or similar new physics experimental signatures at the ILC. 
The sensitivity of the ILC for probing the $Z$-$Z'$ mixing and its
capability to distinguish these two new physics scenarios is
substantially enhanced when the polarization of the initial beams and the produced
$W^\pm$ bosons are considered. 
A model independent analysis of the $Z'$ effects in the process
$e^+e^-\to W^+W^-$ allows to differentiate the full class of vector
$Z'$ models from those with anomalous trilinear gauge couplings,
with one notable exception: the sequential SM (SSM)-like
models can in this process not be distinguished from anomalous gauge couplings.
Results of model
dependent analysis of a specific $Z'$ are expressed in terms
of discovery and identification reaches on the $Z$-$Z'$ mixing angle and
the $Z'$ mass.
\end{abstract}
\pacs{12.60.-i, 12.60.Cn, 14.70.Fm, 29.20.Ej}  
\preprint{DESY 12-016}
\preprint{CERN-PH-TH/2012-077}
\maketitle

\section{Introduction}
The $W^\pm$ boson pair production process
\begin{equation}\label{proc1}
e^+ + e^- \to W^+ + W^-
\end{equation}
is a crucial one for studying the electroweak gauge
symmetry in  $e^+ e^-$ annihilation.
Properties of the weak gauge bosons are closely related to
electroweak symmetry breaking and the structure 
of the gauge sector in general. Thus, detailed examination 
of (\ref{proc1}) at the ILC will both test this sector of the 
standard model (SM) with the highest accuracy and throw light 
on New Physics (NP) that may appear beyond the SM.

In the SM, for zero electron mass, the process (\ref{proc1}) is 
described by the amplitudes mediated by photon and $Z$ boson 
exchange in the $s$-channel and by neutrino exchange in 
the $t$-channel. Therefore, this reaction is particularly sensitive to both the leptonic vertices and the trilinear couplings to $W^+W^-$ of the SM $Z$ and of any new heavy neutral boson that can be exchanged in the $s$-channel. A popular example in this regard, is represented by the $Z^\prime$s envisaged by electroweak scenarios based on spontaneously broken `extended' gauge symmetries, with masses much larger than $M_Z$ and coupling constants different from the SM. The variety of the proposed $Z^\prime$ models is broad. Therefore, rather than attempting an exhaustive analysis, we shall here focus on the phenomenological effects in  reaction (\ref{proc1}) of the so-called $Z^\prime_{\rm SSM}$, $Z^\prime_{E_6}$ and  $Z^\prime_{\rm LR}$ models. Actually, in some sense, we may consider these $Z^\prime$ models as  representative of this New Physics (NP) sector 
\cite{Langacker:2008yv,Rizzo:2006nw,Leike:1996pj,Leike:1998wr,Hewett:1988xc,
Erler:2009jh,Langacker:2009su,Riemann:2005es}.

The direct manifestation of $Z^\prime$s would be the observation of peaks in cross sections at very high energy colliders, this would be possible only for $M_{Z^\prime}$ lying within the kinematical reach of the machine and sufficient luminosity. Indeed, current lower limits on $M_{Z^\prime}$ are obtained 
from direct searches of $Z^\prime$s in Drell-Yan dilepton pair production at the CERN LHC: from the analysis of the 
7 TeV data, the observed bounds at 95\% C. L. range approximately in the interval $1.8 - 2.3~\text{TeV}$, 
depending on the particular $Z^\prime$ model being tested \cite{Chatrchyan:2012it,atlas-dilepton}.   
For too high masses, $Z^\prime$ exchanges can manifest themselves indirectly, {\it via} deviations of cross sections, and in general of the reaction observables, from the SM predictions. Clearly, this kind of searches requires great precision and therefore will be favoured by extremely high collider luminosity, such as will be available at the ILC. Indirect lower bounds on $Z^\prime$  masses from the high precision LEP data at the $Z$ lie in the range $\sim 0.4 - 1.8~\text{TeV}$, depending on the model considered \cite{Erler:2009jh,Langacker:2009su}.

Indirect effects may be quite subtle, as far as the identification of the source of an observed deviation is concerned, because {\it a priori} different NP scenarios may lead to the same or similar experimental signatures. Clearly, then, the discrimination of one NP model (in our case the $Z^\prime$) from other possible ones needs an appropriate strategy for analyzing the data.\footnote{Actually, this should be necessary also in the case of direct discovery, because different NP models may in principle produce the same peaks at the same mass so that, for example, for model identification some angular analyses must be applied, see \cite{Osland:2009tn} and references therein.}     
  
In this paper, we study the indirect effects evidencing the mentioned extra $Z'$ gauge bosons in $W^\pm$ pair production (\ref{proc1}) at the next generation $e^+e^-$ International Linear Collider (ILC), with a center of mass energy $\sqrt s=0.5-1~\text{TeV}$ and typical time-integrated luminosities of ${\cal L}_{\rm int}\sim
0.5-1$~ab$^{-1}$~\cite{:2007sg,Djouadi:2007ik}. At the foreseen, really high luminosity this process should be quite sensitive to the indirect NP effects at a collider with 
$M_Z\ll\sqrt{s}\ll M_{Z^\prime}$ 
\cite{Pankov:1990hq,Pankov:1992cy,Pankov:1994hx,Pankov:1997da,Jung:1999wq,Ananthanarayan:2010bt}, the deviations of cross sections from the SM predictions being expected to increase with $\sqrt{s}$ due to the violation of the SM gauge cancellation among the different contributions. 

Along the lines of the previous discussion, apart from estimating the foreseeable sensitivity of process (\ref{proc1}) 
to the considered $Z^\prime$ models, we will consider the problem of establishing the potential of ILC of distinguishing
the $Z^\prime$ effects, once observed, from the ones 
due to NP competitor models that can lead to analogous physical signatures in the cross section. For the latter, we will choose the models with Anomalous Gauge Couplings (AGC), and compare them with the hypothesis of $Z^\prime$ exchanges. 
In the AGC models, there is no new gauge boson exchange, but the $WW\gamma$, $WWZ$ couplings are modified with respect to the SM values, 
this violates the SM gauge cancellation too and leads to deviations of the process cross sections. AGC couplings are described {\it via} a sum of effective interactions, ordered by dimensionality, and we shall restrict our analysis to the  dimension-six terms which conserve $C$ and $P$
\cite{gounaris,Gounaris:1992kp}. 
 
The baseline configuration of the ILC envisages a very high electron beam polarization (larger than 80\%) that is measurable with high precision. Also positron beam polarization, around 30\%, might be initially obtainable, and 
this polarization could be raised to about 60\% or higher in the ultimate upgrade of the machine. As is well-known, the polarization option represents an asset in order
to enhance the discovery reaches and identification sensitivities
on NP models of any kind \cite{MoortgatPick:2005cw,Osland:2009dp}. This is the case, in 
particular, of $Z^\prime$ exchanges and AGC interactions in process (\ref{proc1}), an obvious example being the suppression of the $\nu$-exchange channel by using right-handed electrons. 
Additional ILC diagnostic ability in $Z^\prime$s and AGC would be provided by measures of polarized $W^+$ and $W^-$ in combination with initial beam polarizations. 

The paper is organized as follows. In Section II, we briefly review the models involving additional $Z^\prime$ bosons and emphasize the role of $Z$-$Z^\prime$ mixing in the process 
(\ref{proc1}). In Section III we give the parametrization of $Z^\prime$ and AGC effects, as well as formulae for helicity amplitudes and cross sections of the
process under consideration. Section IV contains, for illustrative purposes, some plots of the unpolarized and polarized cross sections showing the effect of $Z^\prime$ and of $Z$-$Z^\prime$ mixing. In Section V we present the approach, which allows to obtain the discovery reach
on $Z^\prime$ parameters (actually, on the deviations of the transition amplitudes from the SM) and the obtained numerical results. Section VI includes the results of
both model dependent and model independent analyses of the possibilities to differentiate $Z^\prime$ effects from similar ones caused by AGC. Finally we conclude in Section VII.

\section{$Z^\prime$ models and $Z$-$Z^\prime$ mixing}

The $Z^\prime$ models that will be considered in our
analysis are the following \cite{Langacker:2008yv,Rizzo:2006nw,Leike:1998wr,Hewett:1988xc}:
\begin{itemize}
\item[(i)]
The four possible $U(1)$ $Z'$ scenarios originating from the
spontaneous breaking of the exceptional group $E_{6}$. In this case, two
extra, heavy neutral gauge bosons appear as consequence of the
symmetry breaking and, generally, only the lightest is assumed to
be within reach of the collider. It is defined, in terms of a new
mixing angle $\beta$, by the linear combination
\begin{equation}
Z^\prime=Z^\prime_\chi\cos\beta + Z^\prime_\psi\sin\beta.
\label{beta}
\end{equation}
Specific choices of $\beta$: $\beta=0$; $\beta=\pi/2$;
$\beta=-\arctan{\sqrt{5/3}}$ and $\beta=\arctan{\sqrt{3/5}}$, corresponding to
different $E_{6}$ breaking patterns, define the popular scenarios
$Z^\prime_\chi$, $Z^\prime_\psi$, $Z^\prime_\eta$ and
$Z^\prime_I$, respectively.
\item[(ii)]
The left-right models, originating from the breaking down of an
$SO(10)$ grand-unification symmetry, and where the corresponding
$Z^\prime_{\rm LR}$ couple to a linear combination of
right-handed and $B-L$ neutral currents ($B$ and $L$ being baryon
and lepton numbers, respectively):
\begin{equation}
J^\mu_{\rm LR}= \alpha_{\rm LR}J^\mu_{3R} -\frac{1}{2\alpha_{\rm
LR}}J^\mu_{B-L} \quad {\rm with}\quad \alpha_{\rm LR}=
{\sqrt{\frac{c_W^2}{s_W^2}\,\kappa^2-1}}. \label{left-right}
\end{equation}
Here, $s_W=\sin\theta_W$, $c_W=\sqrt{1-s_W^2}$,
additional parameters are the ratio $\kappa=g_{\rm R}/g_{\rm L}$ of the
$SU(2)_{\rm L,R}$ gauge couplings and $\alpha_{\rm LR}$,
restricted to the range $\sqrt{2/3} \lesssim \alpha_{\rm
LR}\lesssim1.52$. The upper bound corresponds to the so-called
LR-symmetric $Z^\prime_{\rm LRS}$ model with $g_{\rm R}=g_{\rm L}$, while the
lower bound is found to coincide with the $Z^\prime_\chi$ model
introduced above. We will consider the former one, $Z^\prime_{\rm
LRS}$, throughout the paper.
\item[(iii)]
The $Z'_{\rm ALR}$ predicted by the so-called `alternative'
left-right scenario. For the LR model we need not introduce
additional fermions to cancel anomalies. However, in the $E_6$
case a variant of this model (called the Alternative LR model) can
be constructed by altering the embeddings of the SM and introducing exotic
fermions into the ordinary 10 and 5 representations.
\item[(iv)]
The so-called sequential $Z^\prime_\text{SSM}$, where the
couplings to fermions are the same as those of the SM $Z$.
\end{itemize}
Detailed descriptions of these models, as well as the specific
references, can be found, e.~g., in Refs.~\cite{Langacker:2008yv,Rizzo:2006nw,Leike:1998wr,Hewett:1988xc}.

In the extended gauge theories predicting the existence of an extra neutral $Z^\prime$ gauge boson, the mass-squared matrix of the $Z$ and $Z'$ can have non-diagonal entries $\delta M^2$, which are related to the vacuum expectation values of the fields of an extended Higgs sector \cite{Leike:1998wr}:
\begin{equation}\label{massmatrix}
M_{ZZ^\prime}^2 = \left(\begin{matrix} M_Z^2&\delta M^2\\ \delta M^2&M_{Z^\prime}^2
\end{matrix}\right).
\end{equation}
Here, $Z$ and $Z'$ denote the weak gauge boson eigenstates of $SU(2)_L\times U(1)_Y$ and of the extra $U(1)'$, respectively. The mass eigenstates, $Z_1$ and $Z_2$, diagonalizing the matrix (\ref{massmatrix}), are then obtained by the rotation of
the fields $Z$ and $Z^\prime$ by a mixing angle $\phi$:
\begin{eqnarray}
&& Z_1 = Z\cos\phi + Z^\prime\sin\phi\;, \label{z1} \\
&& Z_2 = -Z\sin\phi + Z^\prime\cos\phi\;. \label{z2}
\end{eqnarray}
Here, the mixing angle $\phi$ is expressed in terms of masses 
as:
\begin{equation}
\label{phi} \tan^2\phi={\frac{M_Z^2-M_1^2}{M_2^2-M_Z^2}}\simeq \frac{2 M_Z \Delta
M}{M_2^2}\;,
\end{equation}
where $\Delta M=M_Z-M_1>0$, $M_Z$ is the mass of the $Z_1$-boson in the absence of mixing, i.e., for $\phi=0$. Once we assume the mass $M_1$ to be determined experimentally, the mixing depends on two free parameters, which we identify as $\phi$ and $M_2$. We shall here consider the configuration $M_1\ll\sqrt{s}\ll M_2$.  

The mixing angle $\phi$ will play an important role in our analysis. In general, such mixing effects reflect the underlying gauge symmetry and/or the Higgs sector of the model. 
To a good approximation, for $M_1\ll M_2$, in
specific ``minimal-Higgs models'' \cite{UPR-0476T},
\begin{equation}\label{phi0}
\phi\simeq -s^2_\mathrm{W}\
\frac{\sum_{i}\langle\Phi_i\rangle{}^2I^i_{3L}Q^{\prime}_i}{\sum_{i}\langle\Phi_i\rangle^2(I^i_{3L})^2}={\cal
C}\ {\frac{\displaystyle M^2_1}{\displaystyle M^2_2}}.
\end{equation}
Here $\langle\Phi_i\rangle$ are the Higgs vacuum expectation values
spontaneously breaking the symmetry, and $Q^\prime_i$  are their
charges with respect to the additional $U(1)'$. In addition, in
these models the same Higgs multiplets are responsible for both
generation of mass $M_1$ and for the strength of the $Z$-$Z^\prime$ mixing \cite{Langacker:2008yv}.
Thus ${\cal C}$ is a model-dependent constant. For example, in the
case of $E_6$ superstring-inspired models ${\cal C}$ can be
expressed as \cite{UPR-0476T}
\begin{equation}\label{c}
{\cal C}=4s_\mathrm{W}\left(A-\frac{\sigma-1}{\sigma+1}B\right),
\end{equation}
where $\sigma$ is the ratio of vacuum expectation values squared,
and the constants $A$ and $B$ are determined by the mixing angle $\beta$: $A=\cos\beta / 2\sqrt6$,
$B=\sqrt {10}/12\sin\beta$.

An important property of the models under consideration is that the gauge eigenstate $Z'$ does not couple to the $W^+W^-$ pair since it is neutral under $SU(2)_L$. Therefore the process (\ref{proc1}), and the searched-for deviations of the cross sections from the SM, are sensitive to a $Z^\prime$ only in the case of a non-zero $Z$-$Z'$ mixing. The mixing angle is rather highly constrained, to an upper limit of $\text{a few}\times 10^{-3}$, mainly from LEP measurements at the $Z$~\cite{Erler:2009jh,Langacker:2009su}. The high statistics on $W$-pair production expected at the ILC might in principle allow to probe such small mixing angles effectively.

From (\ref{z1}) and (\ref{z2}), one obtains the
vector and axial-vector couplings of the $Z_1$ and $Z_2$ bosons to fermions:
\begin{eqnarray}
&& v_{1f} = v_f\cos \phi + v_f^\prime \sin \phi\;,\;\;\; a_{1f} = a_f \cos \phi +
a_f^\prime \sin \phi\;, \label{v1} \\
&& v_{2f} = - v_f \sin \phi + v_f^\prime \cos \phi\;,\;\;\;a_{2f} = - a_f \sin \phi +
a_f^\prime \cos \phi, \label{v2}
\end{eqnarray}
with $(v_f,a_f)=(g^f_L\pm g^f_R)/2$, and $(v_f^\prime,a_f^\prime)$ similarly defined in terms of the $Z^\prime$ couplings.
The fermonic $Z^\prime$ couplings can be found in \cite{Langacker:2008yv,Rizzo:2006nw,Leike:1998wr,Hewett:1988xc}.

Analogously, one obtains according to the remarks above:
\begin{eqnarray}
&& g_{WWZ_1}=\cos\phi\;g_{WWZ}\;,  \\
&& g_{WWZ_2}=-\sin\phi\; g_{WWZ}\;,
\end{eqnarray}
where $g_{WWZ}=\cot\theta_W$.

\section{Parameterizations of $Z^\prime$-boson and AGC  effects}

\subsection{$Z'$ boson}

The starting point of our analysis will be the 
amplitude for the process (\ref{proc1}). In the Born
approximation, this can be written as a sum of a $t$-channel and
an $s$-channel component. In the SM
case, the latter will be schematically written  as follows:
\begin{equation}
{\cal M}_s^{(\lambda)}=\left(-\frac{1}{s} +\frac{\cot\theta_W
(v-2\lambda a)}{s-M^2_Z}\right) \times{\cal
G}^{(\lambda)}(s,\theta),\label{amplis}\end{equation} where $s$
and $\theta$ are the total c.m.\ squared energy and $W^-$
production angle. Omitting the fermion subscripts, 
electron vector and axial-vector couplings in
the SM are denoted as
$v=(T_{3,e}-2Q_e\hskip 2pt s_W^2)/2s_Wc_W$ and
$a=T_{3,e}/2s_Wc_W$, respectively, with $T_{3,e}=-1/2$, and  $\lambda$
denoting the electron helicity ($\lambda=\pm1/2$ for
right/left-handed electrons). Finally, ${\cal
G}^{(\lambda)}(s,\theta)$ is a kinematical coefficient, depending
also on the $W^\pm$ helicities. The
explicit form can be found in the literature \cite{gounaris,Gounaris:1992kp} or
derived from the entries of Table~\ref{tab_amplit}, which also
shows the form of the $t$-channel neutrino exchange.

%%%%%%%%%%%%%%%%%%%%%%%%%%%%%%%%%%%%%%%%%%%%%%%%%%%%%%%%%%%%%
\begin{figure}[htb] \vspace{3mm}
\begin{center}
{\includegraphics[angle=90, scale=0.55]{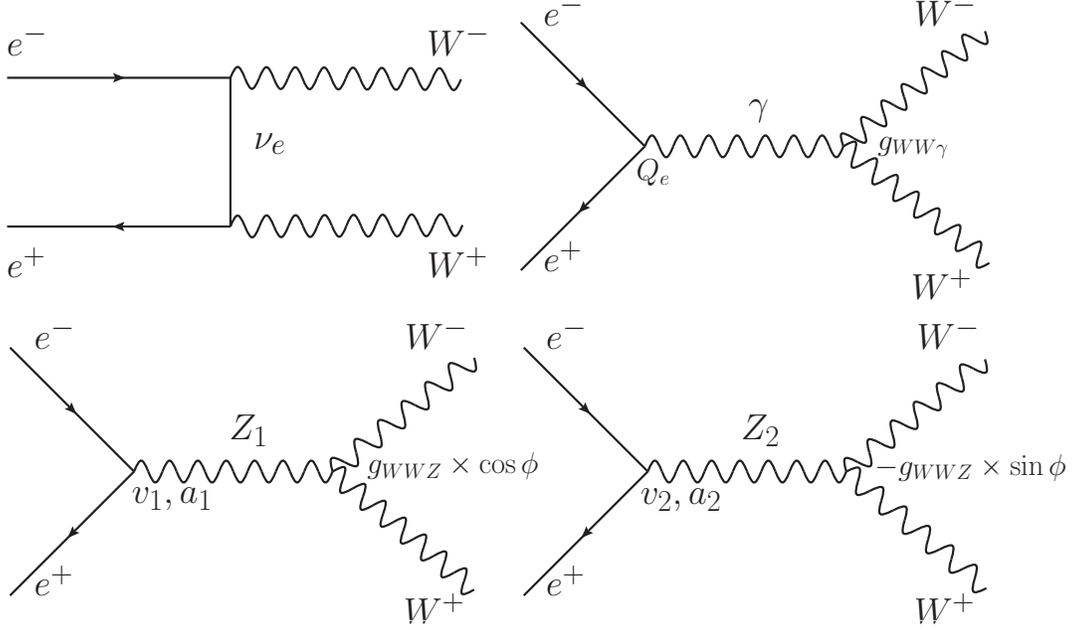}}
\end{center}
\caption{Feynman diagrams for the process $e^- e^+\to
 W^- W^+$ in the Born approximation}\label{fig2}
\end{figure}
%%%%%%%%%%%%%%%%%%%%%%%%%%%%%%%%%%%%%%%%%%%%%%%%%%%%%%%%%%%%%%%

In the extended gauge models the process (\ref{proc1}) is described by the
set of diagrams displayed in Fig.~\ref{fig2}.
The  amplitude  with the extra $Z'$ depicted in
Fig.~\ref{fig2} will be written as:
\begin{equation}{\cal M}_s^{(\lambda)}=\left(-\frac{1}{s}
+\frac{g_{WWZ_1} (v_1-2\lambda a_1)}{s-M_{1}^2} +\frac{g_{WWZ_2}
(v_2-2\lambda a_2)}{s-M_{2}^2}\right) \times{\cal
G}^{(\lambda)}(s,\theta).\label{amplis1}\end{equation}

The contribution of the new heavy neutral gauge boson $Z_2$ to
the amplitude of process (\ref{proc1}) is represented by the fourth
diagram in Fig.~\ref{fig2}. In addition, there are indirect
contributions to the $Z_1$-mediated diagram, represented by
modifications of the electron and three-boson vertices induced
by the $Z$-$Z'$ mixing.

It is convenient to rewrite Eq.~(\ref{amplis1})
in the following form~\cite{Pankov:1997da}:\footnote{Note that $M_Z=M_1+\Delta M$, where $M_1$ refers to the mass eigenstate.}
\begin{equation}
{\cal M}_s^{(\lambda)}=\left(-\frac{g_{WW\gamma}}{s}
+\frac{g_{WWZ}(v-2\lambda a)}{s-M^2_Z}\right) \times{\cal
G}^{(\lambda)}(s,\theta),\label{amplis2}
\end{equation} 
where the `effective' gauge boson couplings $g_{WW\gamma}$ and $g_{WWZ}$ are
defined as:
\begin{equation}
g_{WW\gamma}= 1+\Delta_\gamma=
1+\Delta_\gamma(Z_1)+\Delta_\gamma(Z_2),\label{deltagamma}
\end{equation}
\begin{equation}
g_{WWZ}=\cot\theta_W+\Delta_Z=\cot\theta_W+\Delta_Z(Z_1)+\Delta_Z(Z_2),
\label{deltaz}
\end{equation} with
\begin{equation}\Delta_\gamma(Z_1)=v\hskip 2pt \cot\theta_W\hskip 2pt
\left(\frac{\Delta a}{a}-\frac{\Delta v}{v}\right)\hskip 1pt
\left(1+\Delta\chi\right)\hskip 1pt\chi;\ \
\Delta_\gamma(Z_2)=v\hskip 2pt g_{WWZ_2}\hskip 2pt
\left(\frac{a_2}{a}-\frac{v_2}{v}\right)\hskip 2pt \chi_2,
\label{coupl1}
\end{equation}
\begin{equation}
\Delta_Z(Z_1)=\Delta g_{WWZ}+\cot\theta_W\hskip
1pt\left(\frac{\Delta a}{a} +\Delta\chi\right);\qquad
\Delta_Z(Z_2)= g_{WWZ_2}\hskip 2pt \frac{a_2}{a}\hskip 2pt
\frac{\chi_2} {\chi}.\label{coupl2}\end{equation} In
Eqs.~(\ref{coupl1}) and (\ref{coupl2}) we have introduced the
deviations of the fermionic and trilinear bosonic couplings
$\Delta v=v_1-v$, $\Delta a=a_1-a$ and $\Delta
g_{WWZ}=g_{WWZ_1}-\cot\theta_W$, and the neutral vector boson
propagators (neglecting their widths):
\begin{equation}
\chi (s)=\frac{s}{s-M^2_Z};\qquad \chi_2 (s)=\frac{s}{s-M^2_{2}};
\qquad \Delta\chi (s)\simeq -\frac{2M_Z\Delta M}{s-M^2_Z},\label{chi}
\end{equation}
where $\Delta M=M_Z-M_{1}$ is the $Z$-$Z_1$ mass shift. Because
$W$ pair production is studied sufficiently far away from the
$Z_1$ peak, we can neglect the $Z$ and $Z_{1,2}$ widths in
(\ref{amplis1}) and (\ref{amplis2}).

It should be stressed that, not referring to specific models, the
parametrization (\ref{amplis2})-(\ref{deltaz}) is both general and
useful for phenomenological purposes, in particular to compare
different sources of nonstandard effects contributing finite
deviations (\ref{coupl1}) and (\ref{coupl2}) to the SM
predictions. Note that $\Delta_\gamma$ vanishes as $s\to0$, consistent
with gauge invariance.

We know from current measurements \cite{Erler:2009jh} that $\Delta
M < 100$ MeV. This allows the approximation $\Delta\chi (s)\ll1$.
One can rewrite (\ref{coupl1}) and (\ref{coupl2}) in a
simplified form taking into account the approximation above as
well as the couplings to first order in $\phi$ as:
\begin{equation}
(v_1,\hskip 1pt a_1)\simeq (v+v^\prime\hskip 1pt\phi,\hskip 1pt
a+a^\prime\hskip 1pt\phi) \Rightarrow (\Delta v,\hskip 1pt\Delta
a)\simeq (v^\prime\hskip 1pt \phi, a^\prime\hskip
1pt\phi),\label{v11}\end{equation}
\begin{equation}
(v_2,\hskip 1pt a_2)\simeq(-v\hskip 1pt\phi+v^\prime,\hskip 1pt
-a\hskip 1pt\phi+a^\prime),\label{v22}\end{equation} and
\begin{equation}
g_{WWZ_1}\simeq g_{WWZ}; \qquad\quad
g_{WWZ_2}\simeq -g_{WWZ}\hskip 1pt\phi. \label{g}
\end{equation}
In the case of extended models
considered here, e.g. $E_6$, $v^\prime$ and $a^\prime$ are
explicitly parametrized in terms of the angle $\beta$ which
characterizes the direction of the $Z^\prime$-related extra
$U(1)'$ generator in the $E_6$ group space, and reflects the
pattern of symmetry breaking to $SU(2)_L\times U(1)_Y$
\cite{Langacker:2008yv,Rizzo:2006nw,Leike:1998wr,Hewett:1988xc}:
\begin{equation}
v^\prime=\frac{\cos\beta}{c_W\hskip 1pt\sqrt6};\qquad\quad
a^\prime=\frac{1}{2\hskip 1pt c_W\sqrt 6}\left(\cos\beta+
\sqrt{\frac{5}{3}}\sin\beta\right).\label{vprime}
\end{equation}

Substituting Eqs.~(\ref{v11})--(\ref{g}) into (\ref{coupl1}) and (\ref{coupl2}), one
finds the general form of $\Delta_\gamma$ and $\Delta_Z$:
\begin{equation}
\Delta_\gamma=\phi \cdot v\hskip 2pt \cot\theta_W\hskip 2pt 
\left(\frac{a'}{a}-\frac{v'}{v}\right)\left(1-\frac{\chi_2}{\chi} \right)\chi, 
\label{deltag}
\end{equation}
\begin{equation}
\Delta_Z=\phi \cdot\hskip 2pt \cot\theta_W\hskip 2pt
\frac{a'}{a}\left(1-\frac{\chi_2}{\chi} \right). \label{delta_z}
\end{equation}
Both these quantities have the {\it same} dependence on $\phi$ and $M_2$, via the product $\phi(1-\chi_2/\chi)$. Thus, $\phi$ and $M_2$ can not be separately determined from a measurement of $\Delta_\gamma$ and $\Delta_Z$, only this composite function can be determined.
We also note that for an SSM-type model, the first parenthesis in Eq.~(\ref{deltag}) vanishes,
resulting in $\Delta_\gamma=0$. Thus, these models can not be distinguished from the AGC models, introduced in the next section.
Further, the terms proportional to $\chi_2$ in Eqs. (\ref{deltag})
and (\ref{delta_z}) dominate in the case $\sqrt{s}\approx M_2$ but
will be very small in the case $\sqrt{s}\ll M_2$.

\subsection{Anomalous Gauge Couplings}

As pointed out in the Introduction, a
model with an extra $Z'$ would produce virtual manifestations in
the final $W^+W^-$ channel at the ILC that in principle could
mimic those of a model with AGC, hence of completely different origin.
This is due to the fact that, as shown above, the effects of the extra $Z'$ can be
reabsorbed into a redefinition of the $WWV$ couplings
($V=\gamma,\hskip 2pt Z$). Therefore, the identification of such
an effect, if observed at the ILC, becomes a very important problem \cite{Hagiwara:1986vm}.

%%%%%%%%%%%%%%%%%%%%%%%%%%%%%%%%%%%%%%%%%%%%%%%%%%%%%%%%%%%%%%%%
\begin{figure}[h t b p] \vspace{1mm}
\begin{center}
{\includegraphics[angle=90, scale=0.55]{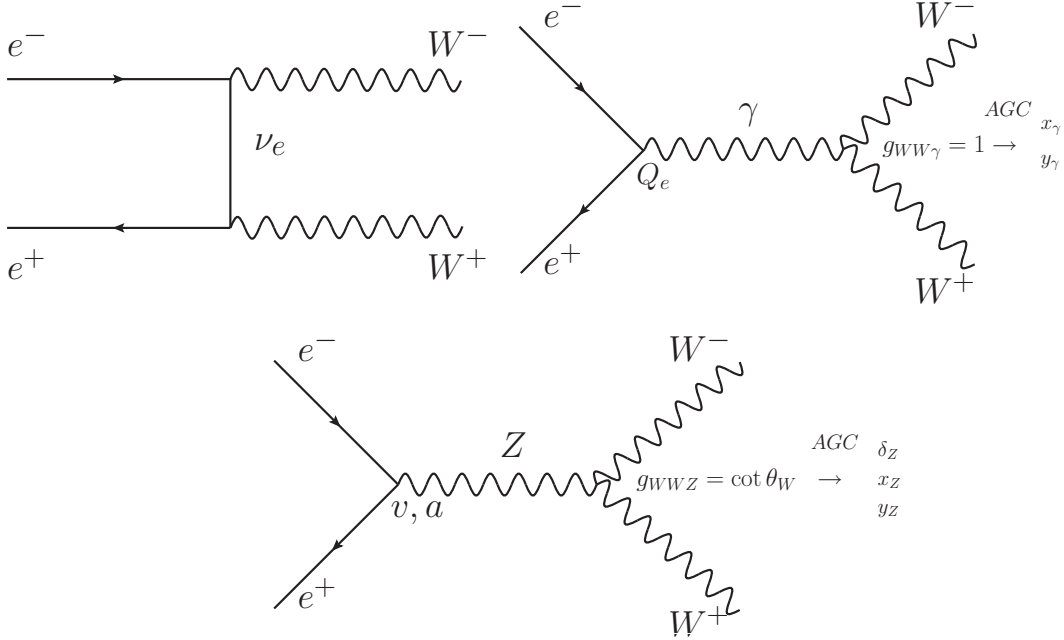}}
\end{center}
\caption{Feynman diagrams for the process $e^+ e^-
\to W^+ W^-$ in the Standard Model and with anomalous trilinear
gauge couplings (AGC). }\label{fig1AGC}
\end{figure}
%%%%%%%%%%%%%%%%%%%%%%%%%%%%%%%%%%%%%%%%%%%%%%%%%%%%%%%%%%%%%%%%

Using the notations of, e.g., Ref.~\cite{gounaris,Gounaris:1992kp}, the relevant
trilinear $WWV$ interaction up to operators of dimension-6, which conserves $U(1)_\text{e.m.}$, $C$ and
$P$, can be written as ($e=\sqrt{4\pi\alpha_{em}}$):
\begin{eqnarray}
{\cal L}_\text{eff}
&=&-ie
\left[A_\mu\left(W^{-\mu\nu}W^+_\nu-
W^{+\mu\nu}W^-_\nu\right)+F_{\mu\nu}W^{+\mu}W^{-\nu}\right]
\nonumber\\
&-& ie\hskip 2pt \left(\cot\theta_W+\delta_Z\right)\hskip 2pt \left[Z_\mu
\left(W^{-\mu\nu}W^+_\nu-W^{+\mu\nu}W^-_\nu\right)+Z_{\mu\nu}W^{+\mu}W^{-\nu}
\right]\nonumber\\
&-&ie\hskip 2pt x_{\gamma}\hskip 2pt F_{\mu\nu}W^{+\mu}W^{-\nu}-
ie\hskip 2pt x_Z\hskip 2pt Z_{\mu\nu}W^{+\mu}W^{-\nu}\nonumber\\
&+&ie\hskip 2pt\frac{y_{\gamma}}{M_W^2}\hskip 2pt
F^{\nu\lambda}W^-_{\lambda\mu} W^{+\mu}_{\ \ \nu}+ie\hskip
2pt\frac{y_Z}{M_W^2}\hskip 2pt Z^{\nu\lambda}
W^-_{\lambda\mu}W^{+\mu}_{\ \ \nu}\hskip 2pt,
\label{lagra}
\end{eqnarray}
where
$W_{\mu\nu}^{\pm}=\partial_{\mu}W_{\nu}^{\pm}-
\partial_{\nu}W_{\mu}^{\pm}$ and ${Z_{\mu\nu}=\partial_{\mu}Z_{\nu}-
\partial_{\nu}Z_{\mu}}$. In the SM at the tree-level, the anomalous couplings
in (\ref{lagra}) vanish: $\delta_Z=x_\gamma=x_Z=y_\gamma=y_Z=0$.

The anomalous gauge couplings are here parametrized in terms of five real independent parameters. This number can be reduced by imposing additional constraints, like local $SU(2)_L\times U(1)_Y$ symmetry, in which case the number would be reduced to three (see for example Tables~2 and 1 of \cite{Bilenky:1993ms} and \cite{Bilenky:1993uy}, respectively).

Current limits reported by the Particle Data Group \cite{Nakamura:2010zzi}, that show the sensitivity to the AGCs attained so far, are roughly of the order of 0.04 for $\delta_Z$, 0.05 for $x_\gamma$, 0.02 for $y_\gamma$, 0.11 for $x_Z$ and 0.12 for $y_Z$. As will be shown in the next sections, at the ILC in the energy and luminosity configuration considered here, sensitivities to deviations  from the SM, hence of indirect New Physics signatures, down to the order of $10^{-3}$ will be reached. This would compare with the expected order of magnitude of the theoretical uncertainty on the SM cross sections after accounting for higher-order corrections to the Born amplitudes of Figs.~\ref{fig2} and \ref{fig1AGC}, formally of order $\alpha_\text{em}$ \cite{Fleischer:1991nw,Beenakker:1994vn}, but that for distributions can reach the size of 10\%, depending on $\sqrt{s}$ \cite{Denner:2005es,Denner:2005fg}.

\subsection{Helicity amplitudes and cross sections}

The general expression for the cross section of process (\ref{proc1}) with
longitudinally polarized electron and positron beams described by the set of diagrams
presented in Fig.~\ref{fig1AGC} can be expressed as
\begin{equation}
\frac{d\sigma}{d\cos\theta}=\frac{1}{4}\left[\left(1+P_L\right)
\left(1-{\bar P}_L\right)\frac{d\sigma^+}{d\cos\theta}+
\left(1-P_L\right)\left(1+{\bar P}_L\right)\frac{d\sigma^-}{d\cos\theta}
\right], \label{longi}\end{equation} where $P_L$ and ${\bar P}_L$
are the actual degrees of electron and positron longitudinal
polarization, respectively, and $\sigma^\pm$ are the cross
sections for purely right-handed ($\lambda = 1/2$) 
and left-handed ($\lambda = -1/2$) electrons.
%with $\lambda^\prime$ the positron helicity.
From Eq.~(\ref{longi}), the cross section for polarized
(unpolarized) electrons and unpolarized positrons corresponds to
$P_L\ne 0$ and ${\bar P_L}=0$ ($P_L ={\bar P_L}=0$).

The polarized cross sections can generally be written as follows:
\begin{equation}
\frac{d\sigma^\pm}{d\hskip 1pt \cos\theta}=\frac{\vert {\bf
p}\vert}{16\pi s\sqrt{s}}\sum_{\tau,\tau'}\vert
F_{\lambda\tau\tau'}(s,\cos\theta)\vert^2.\label{xsection}
\end{equation}
Here, the helicities of the $W^-$ and $W^+$
are denoted by $\tau,\tau'=\pm 1,0$.
Corresponding to the interaction (\ref{lagra}), the helicity amplitudes
$F_{\lambda\tau\tau'}(s,\cos\theta)$ have the structure shown in
Table~\ref{tab_amplit} \cite{gounaris,Gounaris:1992kp} in Appendix~A. In Table~\ref{tab_amplit},
$\beta_W=\sqrt{1-4M_W^2/s}=2p/\sqrt s$, with $p=\vert\boldsymbol{p}\vert$ the c.m.\
momentum of the $W^-$. Furthermore, $s$ and $t$ are the Mandelstam variables, and
$\theta$ the c.m.\ scattering angle, with $t=M_W^2-s(1-\beta\cos\theta)/2$.
For comparison, we also show in Appendix~A the corresponding helicity amplitudes for the case of a $Z^\prime$.

We define the differential cross sections for correlated spins of the produced
$W^-$ and $W^+$,
\begin{equation}
\frac{d\sigma(W^+_LW^-_L)}{d\cos\theta},\qquad
\frac{d\sigma(W^+_TW^-_T)}{d\cos\theta},\qquad
\frac{d\sigma(W^+_TW^-_L+W^+_LW^-_T)}{d\cos\theta},
\end{equation}
which correspond to the production of  two longitudinally
($\tau=\tau'=0$), two transversely ($\tau=\pm\tau'$;
$\tau,\tau'=\pm 1$) and one longitudinally plus one transversely
($\tau=0$, $\tau'=\pm 1$ etc.) polarized vector bosons,
respectively.

\section{$Z^\prime$ Illustrations}

\begin{figure}[htb]
\includegraphics[scale=0.65]{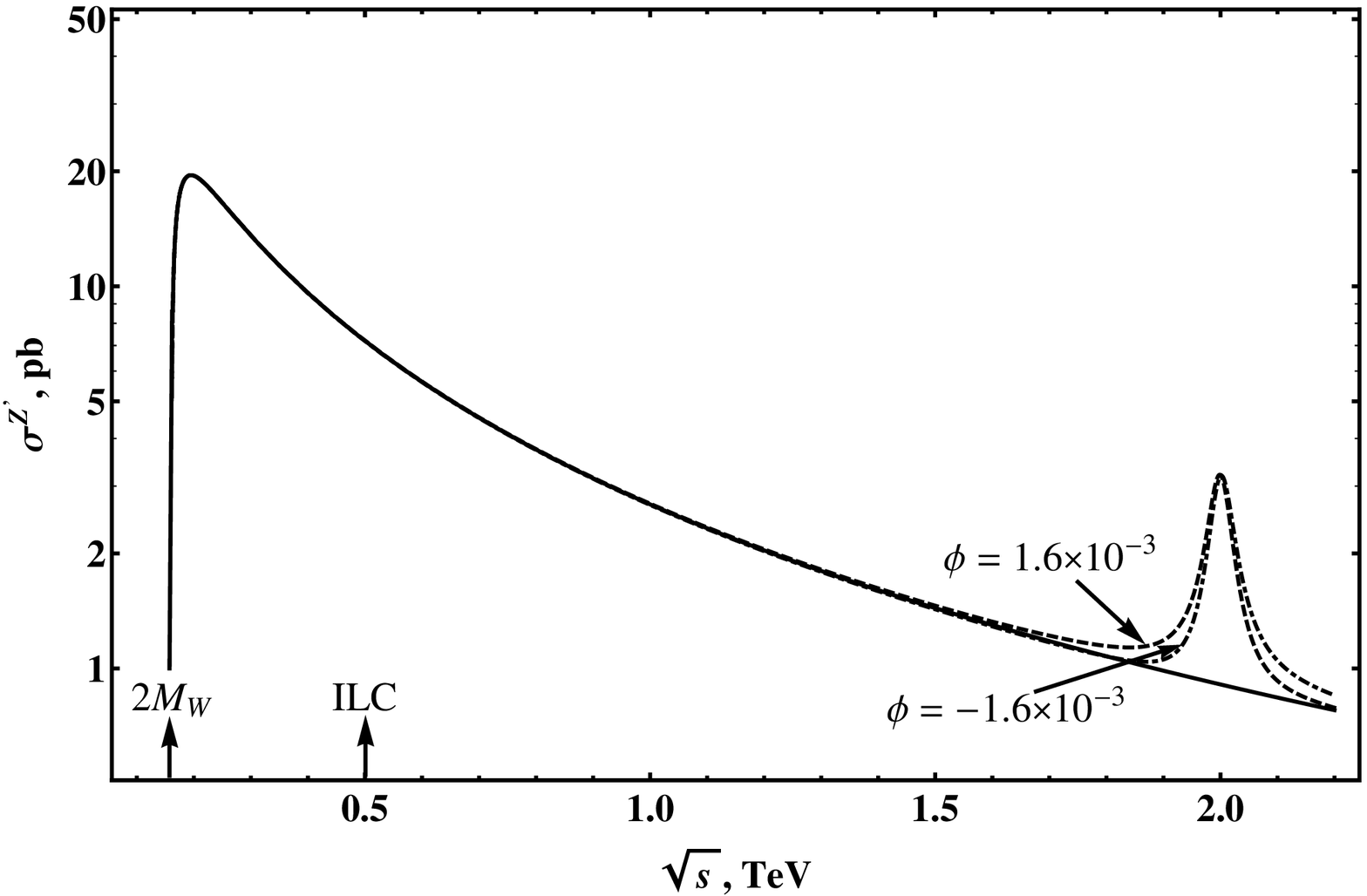}
\vspace{5mm}\hspace*{-5mm}
\includegraphics[scale=0.65]{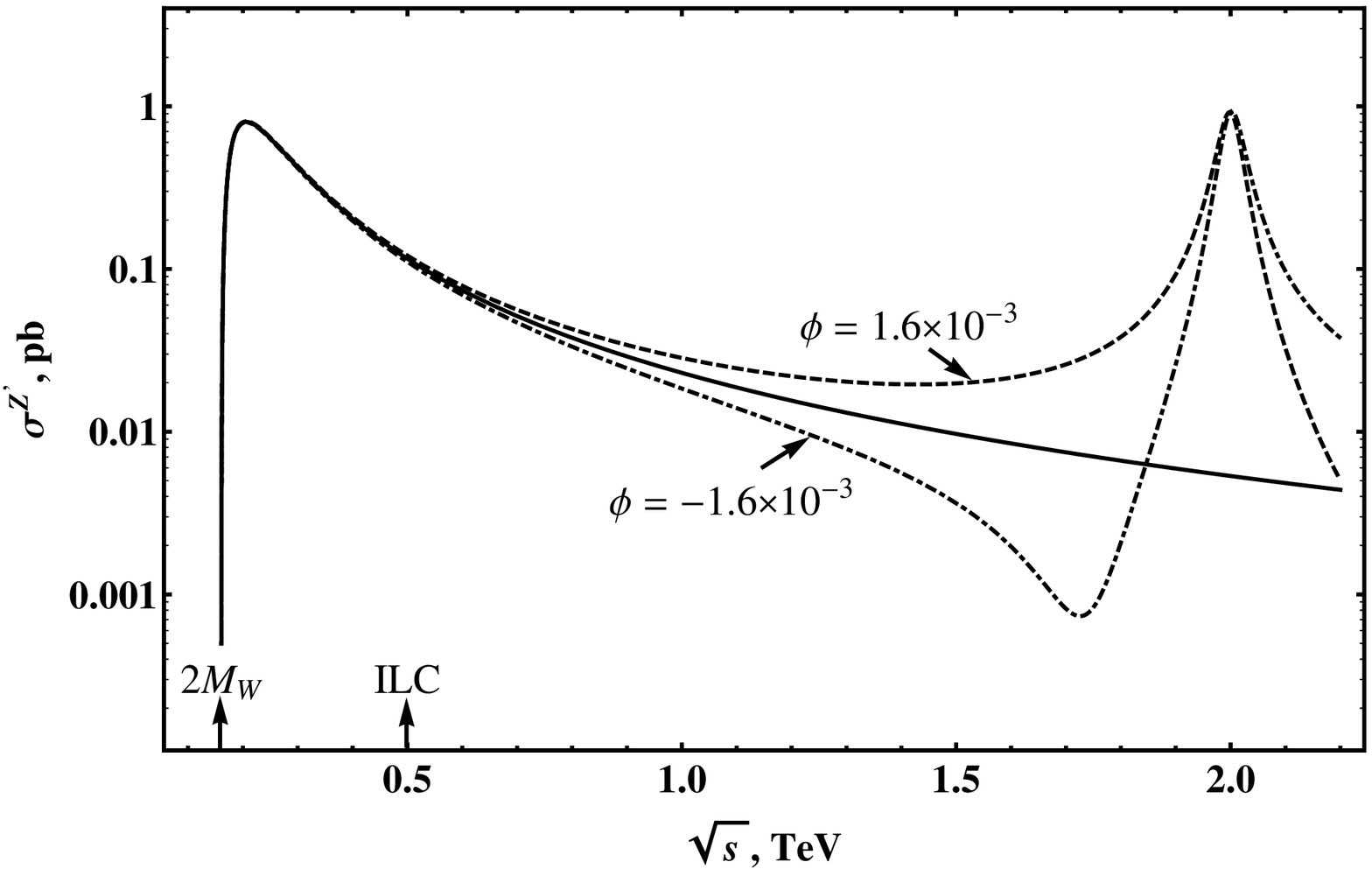}
\vspace{-3mm} 
\caption{Top panel: Unpolarized total cross section for the
process $e^+e^-\rightarrow W^+W^-$ for $Z'_\chi$ from $E_6$. Bottom panel: Polarized
total cross section. Solid lines correspond to the SM case. Dashed (dash-dotted) lines
correspond to a $Z^\prime$ model with $\phi = 1.6\cdot10^{-3}$ ($\phi = -1.6\cdot
10^{-3}$), $\Gamma_2=0.025\times M_2$ and $M_{2} = 2~\text{TeV}$. } \label{fig2-3}
\end{figure}

For illustrative purposes, the energy behavior of the total unpolarized cross section
for the process $e^+e^-\rightarrow W^+W^-$ is shown in Fig.~\ref{fig2-3} (top panel)
for the SM (extrapolated to 2~TeV) as well as for the case of an additional $Z'_\chi$ originated from $E_6$ at
mixing angle $\phi = \pm\ 1.6\cdot 10^{-3}$ and $M_{Z^\prime} = 2~\text{TeV}$. In the
lower panel we show the corresponding cross section for right-handed electrons
($P_L=1$). The deviation of the cross sections from the SM prediction caused by the
$Z'$ boson at the planned ILC energy of $\sqrt{s}=0.5~\text{TeV}$ is most pronounced
for the latter (polarized) case while the cross section is lower than that for
unpolarized beams. The main reason for this is the removal of the neutrino exchange in
the $t$-channel. Such a removal is indispensable for evidencing the $Z^\prime$-exchange effect 
through $Z$--$Z^\prime$ mixing in the process (\ref{proc1}). 
The complete removal of the neutrino exchange contribution depends of course on
having pure electron polarization.
In both cases experimental constraints on the $W^-$ scattering angle
($|\cos\theta|\leq 0.98$) were imposed.

%%%%%%%%%%%%%%%%%%%%%%%%%%%%%%%%%%%%%%%%%%%%%%%%%%%%%%%%
%%%%%%%%%%%%%%%%%%%%%%%%%%%%%%%%%%%%%%%%%%%%%%%%%%%%%%%%%%%%

%%%%%%%%%%%%%%%%%%%%%%%%%%%%%%%%%%%%%%%%%%%%%%%%%%%%%%%%%%%%%%
\begin{figure}[htb]
\includegraphics[scale=0.70]{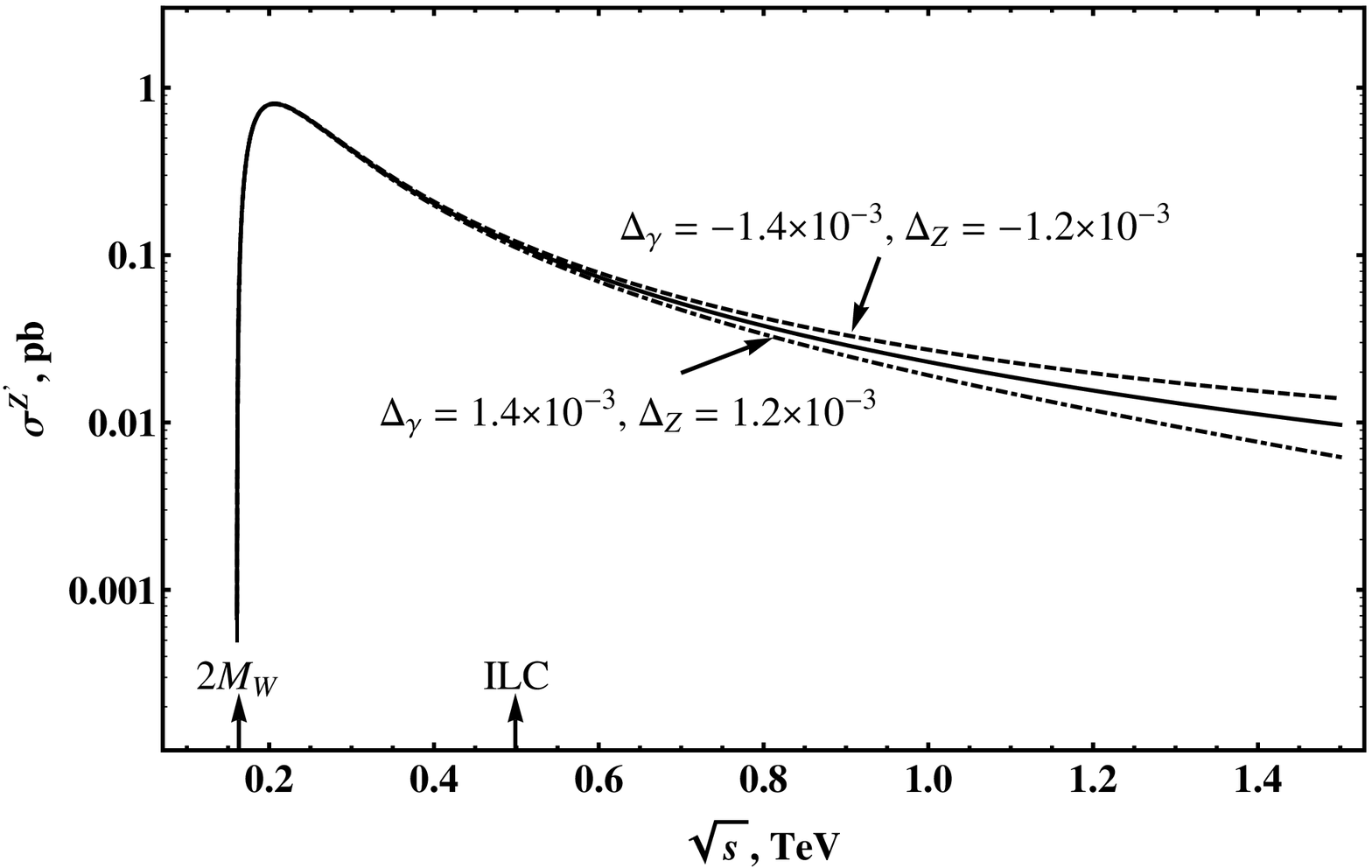}
\hspace{0mm}\vspace{5mm}
\includegraphics[scale=0.69]{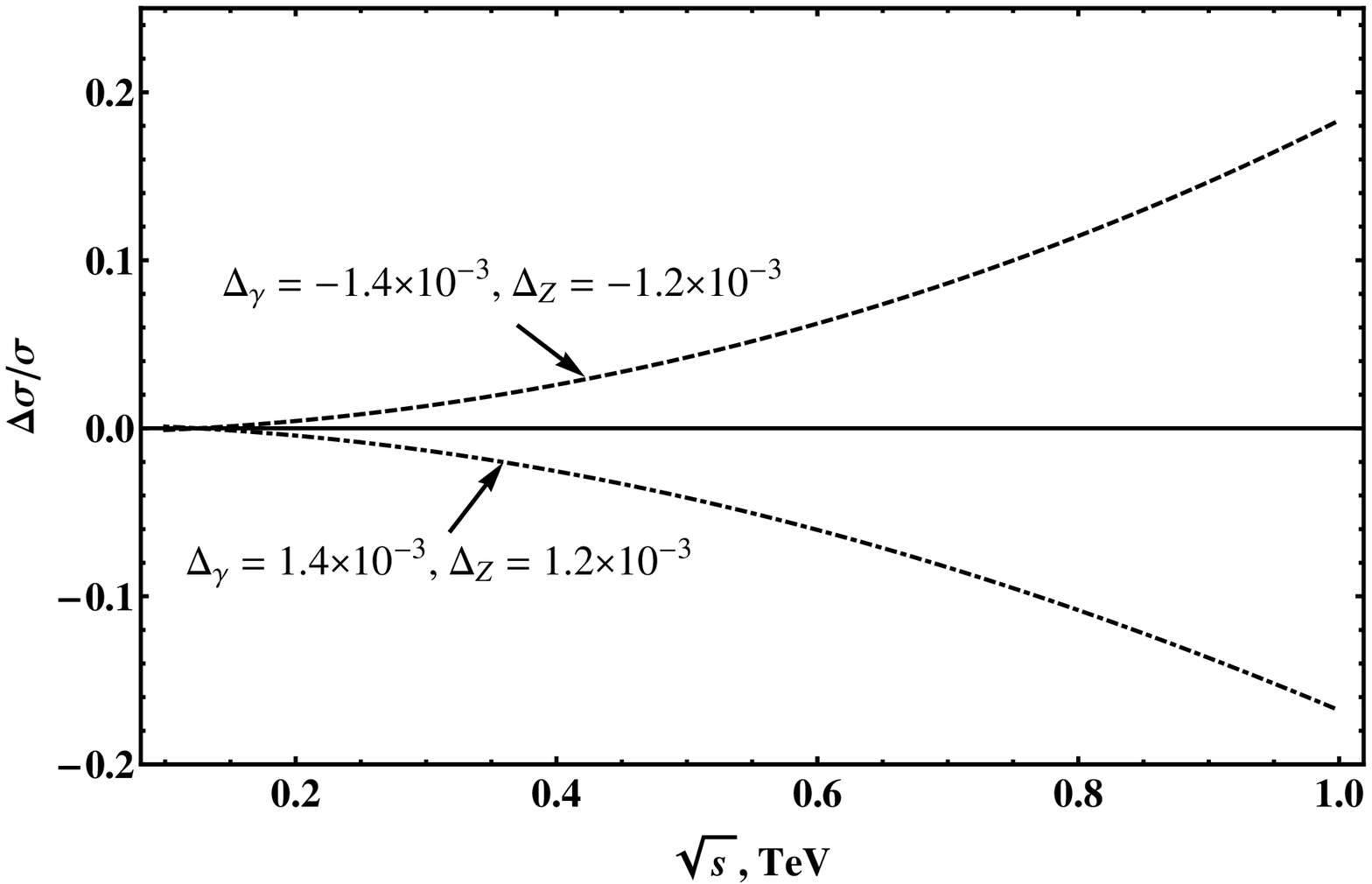}
\hspace{0mm}\vspace{-3mm} \caption{Top panel: Polarized total cross section for the
process $e^+e^-\rightarrow W^+W^-$ as a function of $\sqrt{s}$ with perfectly
polarized electrons ($P_L=1$) and unpolarized final states. Solid line corresponds to
the SM. Contribution to the cross section caused by $Z^\prime$ is determined by
different sets of parameters
 $(\Delta_\gamma,\Delta_Z)=(1.4 \cdot 10^{-3},1.2 \cdot 10^{-3})$ and $(-1.4 \cdot 10^{-3},-1.2 \cdot 10^{-3})$.
 Bottom panel: Relative deviation of
the polarized total cross section from the SM prediction, $\Delta\sigma/\sigma =
(\sigma^{Z^\prime}-\sigma^{SM})/\sigma^{SM}$. }
 \label{fig5-6}
\end{figure}
%%%%%%%%%%%%%%%%%%%%%%%%%%%%%%%%%%%%%%%%%%%%%%%%%%%%%%%%%%%%%%%%

The effects of the $Z'$ boson shown in Fig.~\ref{fig2-3} were parametrized by the mass
$M_{Z^\prime}$ and the $Z$-$Z'$ mixing angle $\phi$ while those behaviors  and their
relative deviations shown in Fig.~\ref{fig5-6}, are parametrized by the effective
parameters $(\Delta_\gamma,\Delta_Z$), defined in Eqs.~(\ref{deltag}) and (\ref{delta_z}) 
for the same values of $\phi$ and $M_{Z^\prime}$. Rather steep energy behavior  of relative
deviations of the cross sections can be appreciated from Fig.~\ref{fig5-6}.

As was mentioned in the Introduction, the process (\ref{proc1}) is
sensitive to a $Z'$ in the case of non-zero $Z$-$Z'$ mixing. The
individual (interference) contributions to the cross section of
process (\ref{proc1}) rise proportional to $s$. In the SM, the sum over
all contributions to the total cross section results in its proper
energy dependence that scales like $\log{s}/s$ in the limit when
$2M_W\ll\sqrt{s}\ll M_2$ due to a delicate gauge cancellation. In the
case of a non-zero $Z$-$Z'$ mixing, the couplings of the $Z_1$
differ from those of the SM predictions for  $Z$. Then, the gauge
cancellation occurring in the SM is destroyed, leading to an
enhancement of new physics effects at high energies, though well
below $M_2$. Unitarity is restored only at energies
$\sqrt{s}\gg M_2$ independently of details of the extended gauge
group.

%%%%%%%%%%%%%%%%%%%%%%%%%%%%%%%%%%%%%%%%%%%%%%%%%%%%%%%%%%%%%%%%
\section{Discovery reach on  $Z'$ parameters}
\label{sect:discovery-zprime}
%%%%%%%%%%%%%%%%%%%%%%%%%%%%%%%%%%%%%%%%%%%%%%%%%%%%%%%%%%%%%%%%

The sensitivity of the polarized differential cross sections  to $\Delta_\gamma$ and
$\Delta_Z$ is assessed numerically by dividing the angular range
$\vert\cos\theta\vert\leq 0.98$ into 10 equal bins, and defining a $\chi^2$ function
in terms of the expected number of events $N(i)$ in each bin for a given combination of beam polarizations:
\begin{equation}
\chi^{2}=\chi^2(\sqrt{s},\Delta_\gamma,\Delta_Z)
=\sum_{\{P_L,\,\bar{P}_L\}}\sum^\text{bins}_i\left[\frac{N_{\text{SM}+Z^\prime}(i)-N_\text{SM}(i)}
{\delta N_\text{SM}(i)}\right]^2,\label{chi2}\end{equation} where
$N(i)=\Lumint\,\sigma_i\,\varepsilon_W$ with $\Lumint$ the time-integrated luminosity.
Furthermore,
\begin{equation}
\sigma_i=\sigma(z_i,z_{i+1})= \int
\limits_{z_i}^{z_{i+1}}\left(\frac{d\sigma}{dz}\right)dz,
\label{sigmai}
\end{equation}
where $z=\cos\theta$ and polarization indices have been suppressed.
Also, $\varepsilon_W$ is the
efficiency for $W^+W^-$ reconstruction, for which we take the
channel of lepton pairs ($e\nu+\mu\nu$) plus two hadronic jets,
giving $\varepsilon_W\simeq 0.3$ basically from the relevant
branching ratios.  The  procedure outlined above is followed to
evaluate both $N_\text{SM}(i)$ and $N_{\text{SM}+Z^\prime}(i)$.

The uncertainty on the number of events $\delta N_\text{SM}(i)$ combines both
statistical and systematic errors where the statistical component is determined by
$\delta N^\text{stat}_\text{SM}(i)= \sqrt{N_\text{SM}(i)}$. Concerning systematic
uncertainties, an important source is represented by the uncertainty on beam
polarizations, for which we assume $\delta P_L/P_L= \delta {\bar P_L}/{\bar P_L}=
0.5\%$ with the ``standard'' envisaged values $\vert P_L\vert=0.8$ and $\vert
\bar{P}_L\vert=0.5$ \cite{:2007sg,Djouadi:2007ik,MoortgatPick:2005cw}. As for the
time-integrated luminosity, for simplicity we assume it to be equally distributed
between the different polarization configurations. Another source of systematic
uncertainty originates from the efficiency of reconstruction of $W^\pm$ pairs which we
assume to be $\delta\varepsilon_W/\varepsilon_W=0.5\%$. Also, in our numerical
analysis to evaluate the sensitivity of the differential distribution to model
parameters we include initial-state QED corrections to on-shell $W^\pm$ pair
production in the flux function approach \cite{Beenakker:1991jk,Beenakker:1990sf} that assures a good
approximation within the expected accuracy of the data.

%%%%%%%%%%%%%%%%%%%%%%%%%%%%%%%%%%%%%%%%%%%%%%%%%%%%%%%%%%%%%
\begin{figure}[htb]
\includegraphics[scale=0.69]{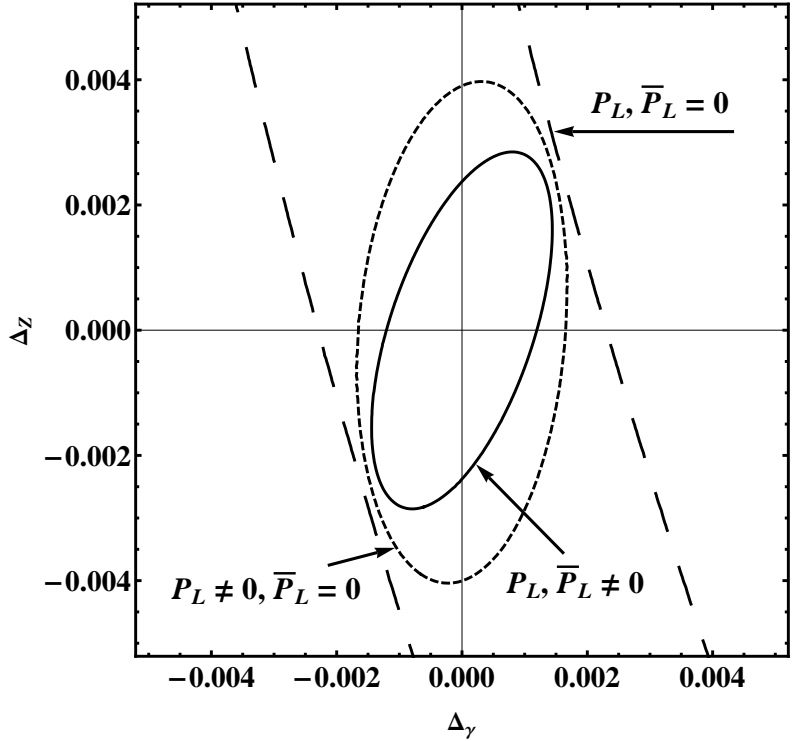}
\vspace{-3mm}
\caption{Discovery reach (see Eq.~(\ref{Eq:chi_sq})) at 95\% CL %($1 - p = 0.95$)
on the $Z^\prime$ parameters $\Delta_\gamma,\Delta_Z$ obtained from polarized differential
cross sections at different sets of
 polarization: $P_{L}=\pm 0.8,\;\bar{P}_{L}=\mp 0.5$ (solid line),
 $P_{L}=\pm 0.8,\;\bar{P}_{L}= 0$ (short-dashed line),
unpolarized beams $P_{L}= 0,\;\bar{P}_{L}= 0$
 (long-dashed line), $\sqrt{s}=0.5~\text{TeV}$ and $\Lumint=500~\text{fb}^{-1}$.}\label{fig7}
\end{figure}
%%%%%%%%%%%%%%%%%%%%%%%%%%%%%%%%%%%%%%%%%%%%%%%%%%%%%%%%%%%%%%%%

As a criterion to derive the constraints on the coupling constants in the case where
no deviations from the SM were observed within the foreseeable uncertainties on the measurable cross sections, we impose that 
\begin{equation} \label{Eq:chi_sq}
\chi^2\leq \chi^2_{\mathrm{min}} + \chi^2_\text{CL}, 
\end{equation}
where $\chi^2_\text{CL}$ is a number that
specifies the chosen confidence level, $\chi^2_{\mathrm{min}}$ is the minimal value of
the $\chi^2$ function. With two independent parameters in Eqs.~(\ref{deltagamma}) and
(\ref{deltaz}), the $95\%$ CL is obtained by choosing $\chi^2_\text{CL}=5.99$.

%%%%%%%%%%%%%%%%%%%%%%%%%%%%%%%%%%%%%%%%%%%%%%%%%%%%%%%%%%%%%%%%
\begin{figure}[!htb]
\includegraphics[scale=0.6]{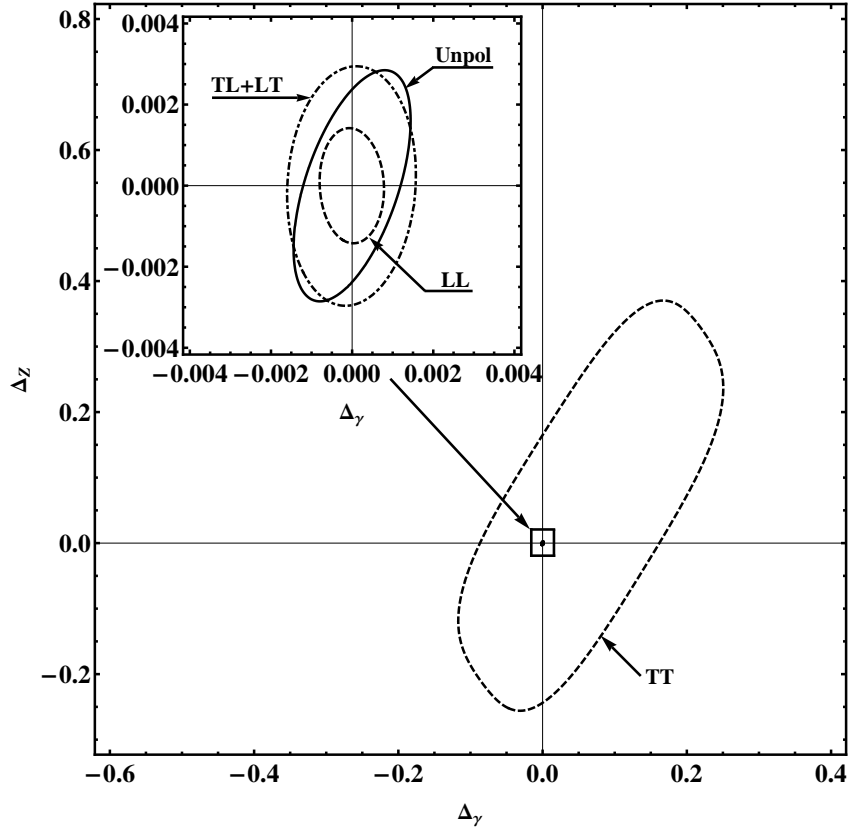}
\caption{Discovery reach %($1 - p = 0.95$)
on the $Z^\prime$ parameters $\Delta_\gamma,\Delta_Z$ from the cross section with polarized
beams $P_{L}=\pm 0.8,\;\bar{P}_{L}=\mp 0.5$ and different sets of $W^\pm$
polarizations.  Here, $\sqrt{s}=0.5~\text{TeV}$ and $\Lumint=500~\text{fb}^{-1}$.}\label{fig8}
\end{figure}
%%%%%%%%%%%%%%%%%%%%%%%%%%%%%%%%%%%%%%%%%%%%%%%%%%%%%%%%%%%%%%%%

From the numerical procedure outlined above, we obtain the allowed regions in
$\Delta_\gamma$ and $\Delta_Z$ determined from the differential  polarized cross
sections with different sets of polarization (as well as from the unpolarized process
(\ref{proc1})) depicted in Fig.~\ref{fig7}, where $\Lumint=500~\text{fb}^{-1}$ has
been taken \cite{:2007sg,Djouadi:2007ik,MoortgatPick:2005cw}. 
According to the condition (\ref{Eq:chi_sq}), the values of $\Delta_\gamma$ and $\Delta_Z$
for which $Z'$s can be discovered at the ILC is represented by the region
external to the ellipse. The same is true for the AGC
model except that, having assumed no renormalization of
the residue of the photon pole exchange ($\delta_\gamma=0$), in
this case $\Delta_\gamma$ will be proportional to $s$ times the
coefficients $x_\gamma$ or $y_\gamma$ of Eq.~(\ref{lagra}), and $\Delta_Z$ to
a combination of the coefficients $\delta_Z$, $x_Z$ and $y_Z$ (see
Table~\ref{tab_amplit}).
The role of initial beam polarization is seen to be essential
in order to set meaningful finite bounds on the parameters.

%%%%%%%%%%%%%%%%%%%%%%%%%%%%%%%%%%%%%%%%%%%%%%%%%%%%%%%%%%%%%%%%
\begin{figure}[h t p b]
\includegraphics[scale=0.57]{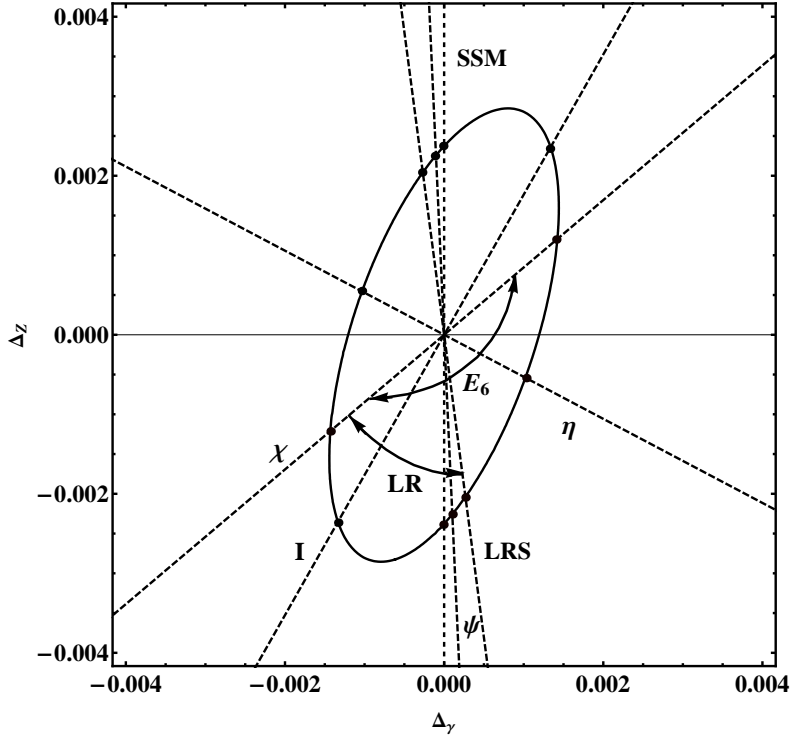}
\caption{Discovery reach (95\% C.L.) %($1 - p = 0.95$)
on $Z^\prime$ parameters  ($\Delta_\gamma$, $\Delta_Z$) obtained from differential
polarized cross sections with ($P_{L}=\pm 0.8,\;\bar{P}_{L}=\mp 0.5$).
 Dashed straight lines correspond to specific extended gauge models
($\chi$, $\psi$, $\eta$, $I$ and LRS) according to Eq.~(\ref{relation}).
The segments of the ellipse correspond to the
whole classes of $E_6$ and LR-models, respectively. 
Here, $\sqrt{s}=0.5~\text{TeV}$ and $\Lumint=500~\text{fb}^{-1}$.}\label{fig8a}
\end{figure}
%%%%%%%%%%%%%%%%%%%%%%%%%%%%%%%%%%%%%%%%%%%%%%%%%%%%%%%%%%%%%%%%

Analogous to Fig.~\ref{fig7}, the discovery reach %($1 - p = 0.95$)
on the parameters $\Delta_\gamma,\Delta_Z$ from the cross section with polarized beams
$P_{L}=\pm 0.8,\;\bar{P}_{L}=\mp 0.5$ and different sets of $W^\pm$ polarizations is
depicted in Fig.~\ref{fig8} which demonstrates that ${d\sigma(W^+_LW^-_L)}/{dz}$ is
most sensitive to the parameters $\Delta_\gamma,\Delta_Z$ while
${d\sigma(W^+_TW^-_T)}/{dz}$ has the lowest sensitivity to those parameters.
The reason for the lower sensitivity in the $TT$ case is that for $s\gg M_Z^2$,
the NP contributions to these amplitudes
only interfere with a sub-dominant part of the SM amplitude \cite{Bilenky:1993ms}.

As regards the NP scenarios of interest here, one may remark that constraints on
$\Delta_\gamma$ and $\Delta_Z$ of Eqs.~(\ref{deltagamma}) and (\ref{deltaz}) 
(for the example of $Z'$s),
are model-independent in the sense that they constrain the whole
class of $Z'$ models considered. They may turn into constraints on
the parameters of specific $Z'$ models by replacing expressions (\ref{coupl1}) and
(\ref{coupl2}). Specializing to those models, one can notice the important
linear relation characterizing the deviations from the SM:
\begin{equation}
\Delta_Z=\Delta_\gamma\cdot\hskip 1pt\frac{1}{v\hskip 1pt\chi}\hskip 1pt
\frac{(a^\prime/a)}{(a^\prime/a)-(v^\prime/v)}, \label{relation}
\end{equation}
where $v$ and $a$ refer to vector and axial-vector couplings.
This relation is rather unique, and depends neither on
$\phi$ nor on $M_{2}$,  only on ratios of the electron couplings
with the $Z$ and $Z'$ bosons. 

In Fig.~\ref{fig8a} we depict, as an illustration, the cases corresponding to the
models  denoted  $\chi$, $\psi$, $\eta$ and $I$ originated from $E_6$ as well as the LR
symmetric model (LRS).  The model independent bound on $\Delta_\gamma$ and $\Delta_Z$
can be converted into limits on the $Z$-$Z'$ mixing angle $\phi$ and mass $M_2$
for any specific $Z'$ model. These model dependent constraints will be presented in
the next section along with identification reaches. For fixed $\phi$ and $M_2$, every model is
represented by a point in the ($\Delta_\gamma$, $\Delta_Z$) parameter plane.
The discovery regions in the $\Delta_\gamma$--$\Delta_Z$ plot at the ILC are
represented by the straight segments lying outside the ellipse.
If one varies the mixing angle $\phi$, the point representative of the
specific $Z'$ model moves along the corresponding line.
The intercept of the lines with the elliptic contour,
once translated to $\phi$ and $M_2$, determine the constraint on these
two parameters relevant to $Z$-$Z'$ mixing for the individual models.

Also, one can determine the region in
the ($\Delta_\gamma,\Delta_Z$) plane relevant to constraining the full
class of $E_6$ (and LR) $Z'$ models obtained by varying the parameters
$\cos{\beta}$ and $\alpha_\text{LR}$ of Eqs.~(\ref{beta}) and (\ref{left-right}) within their full
allowed ranges. The corresponding discovery region at the ILC for
that class of models is the one delimited by the arcs of ellipse
indicated in Fig.~\ref{fig8a}.

%%%%%%%%%%%%%%%%%%%%%%%%%%%%%%%%%%%%%%%%%%%%%%%%%%%%%%%%%%%%%%%%
\section{Identification of $Z'$ vs AGC}
\subsection{Model independent analysis}
%%%%%%%%%%%%%%%%%%%%%%%%%%%%%%%%%%%%%%%%%%%%%%%%%%%%%%%%%%%%%%%%

We will here discuss how one can differentiate various $Z^\prime$ models from  similar
effects caused by anomalous gauge couplings, following the procedure employed in
Refs.~\cite{Pankov:2005kd,Osland:2009dp}. 
The philosophy is as follows: A particular $Z^\prime$ model will be considered identified, if the measured values of $\Delta_\gamma$ and $\Delta_Z$ are statistically different from values corresponding to other $Z^\prime$ models (for a discussion, see Ref.~\cite{Osland:2009dp}), and also different from ranges of $(\Delta_\gamma,\Delta_Z)$ that can be populated by AGC models. Clearly, at least one of these parameters must exceed some minimal value.

Let us assume the data to be
consistent with one of the $Z^\prime$ models and call it the ``true'' model. It has some non-zero values of the parameters $\Delta_\gamma,\Delta_Z$. We want to assess the
level at which this ``true'' model is distinguishable from the AGC models, that can
compete with it as sources of the assumed deviations of the cross section from the SM and we call them ``tested''
models, for any values of the corresponding AGC parameters. 
We assume for simplicity that all AGC parameters are zero, except the one whose values
are probed.

We start by considering as a ``tested'' AGC model
that with a value of $x_\gamma$ to be scanned over.
To that purpose, we can define a ``distance'' between the chosen ``true'' model and the ``tested'' AGC
model(s) by means of a ${\chi}^2$ function analogous to
Eq.~(\ref{chi2}) as
\begin{equation}
\chi^{2}=\sum_{\{P_L,\,\bar{P}_L\}}\sum^\text{bins}_i\left[\frac{N_{Z'}(i)-N_\text{AGC}(i)}
{\delta N_{Z'}(i)}\right]^2,\label{chi2id}
\end{equation}
with $\delta N_{Z'}(i)$ defined in the same way as $\delta N_\text{SM}(i)$ but, in
this case, the statistical uncertainty refers to the $Z'$ model and therefore depends
on the relevant, particular, values of $\Delta_\gamma$ and $\Delta_Z$.

On the basis of such $\chi^2$ we can study whether these ``tested'' models can be excluded or not to a given confidence level (which we assume to be 95\%), once the considered $Z'$ model (defined in terms of $\Delta_\gamma$, $\Delta_Z$) has been assumed as ``true''.
In our explicit example, we want to determine the range in $x_\gamma$ for which there is ``confusion'' of deviations from the SM cross sections between the selected ``true'' $Z^\prime$ model and the AGC one, by imposing the condition, similar to  Eq.~(\ref{Eq:chi_sq}). Then we scan all values of $\Delta_\gamma$, $\Delta_Z$ allowed by the $Z^\prime$ models down to their discovery reach, and determine by iteration in this procedure the general confusion region between the class of $Z^\prime$ models considered here and the AGC model with $x_\gamma\neq 0$.

Besides the dependence on the c.m.\ energy $\sqrt{s}$, the $\chi^2$ function defined above can be considered a function of three independent variables, $\Delta_\gamma$ and $\Delta_Z$ from the $Z^\prime$ model, and, in our starting example, the parameter $x_\gamma$ of the AGC scenario.
The contours of the confusion regions, at given $\sqrt{s}$, are thus defined by the region {\it inside} of which (in the $\Delta_\gamma$-$\Delta_Z$ space)
\begin{equation} \label{Eq:contour}
\chi^2(\Delta_\gamma,\Delta_Z,x_\gamma)
=\chi^2_{\mathrm{min}} + \chi^2_\text{CL},
\end{equation}
for any value of $x_\gamma$ compatible with experimental limits.

%%%%%%%%%%%%%%%%%%%%%%%%%%%%%%%%%%%%%%%%%%%%%%%%%%%%%%
\begin{figure}[htb]
\includegraphics[scale=0.64]{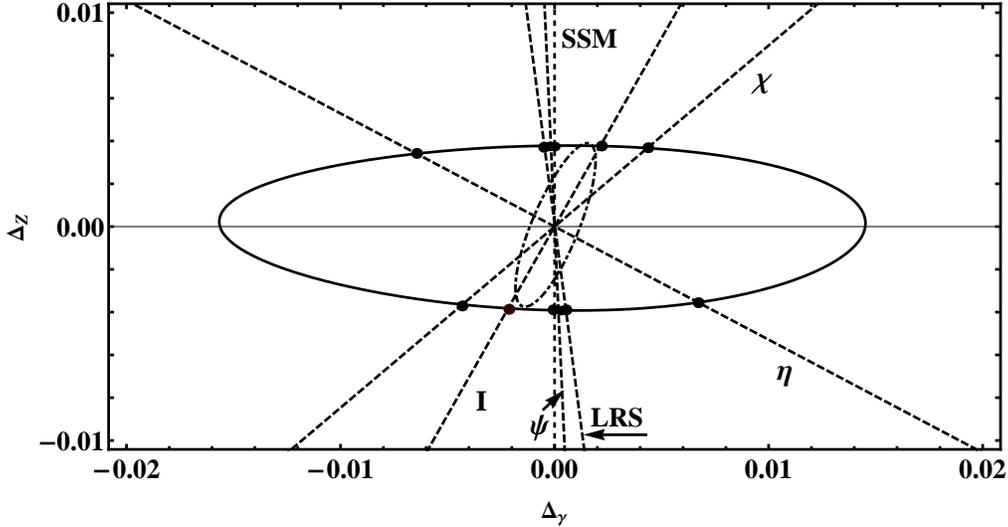}
\vspace{-5mm}
\caption{The outer ellipse (solid) shows the confusion region (95\% C.L., see Eq.~(\ref{Eq:contour}))
%($1 - p = 0.95$)
in the parameter plane ($\Delta_{\gamma},\Delta_Z$), outside of which a generic $Z'$ model can be identified against an 
AGC model with non-vanishing parameter $x_\gamma$. Polarized cross
section with $P_{L}=\pm 0.8$ and $\bar{P}_{L}=\mp 0.5$ are assumed. The dashed inner ellipse reproduces the discovery reach on the $Z^\prime$ of Fig.~\ref{fig8a}, corresponding to $x_\gamma=0$ in Eq.~(\ref{Eq:contour}), where the AGC model coincides with the SM. The dashed straight lines correspond to specific extended gauge models
($\chi$, $\psi$, $\eta$, I and LRS). Here, $\sqrt{s}=0.5~\text{TeV}$ and $\Lumint=500~\text{fb}^{-1}$.
} \label{fig9}
\end{figure}
%%%%%%%%%%%%%%%%%%%%%%%%%%%%%%%%%%%%%%%%%%%%%%%%%%%%%%%%%

In Fig.~\ref{fig9} we show the region of confusion in the $Z'$ parameter plane
($\Delta_\gamma, \Delta_Z$), outside of which the $Z^\prime$ model can be identified at the 95\% C.L.
against the AGC model for any value of the parameter $x_\gamma$. It is
obtained from the polarized cross section with $P_{L}=\pm
0.8$ and $\bar{P}_{L}=\mp 0.5$ using the algorithm outlined above. 
Also, note that the
inner dash-dotted ellipse in Fig.~\ref{fig9} delimits the discovery reach on
$Z'$ parameters.

The graphical representation of the region of confusion presented in Fig.~\ref{fig9}
is straightforward. Equation (\ref{Eq:contour}) defines a three-dimensional surface enclosing a volume in
the ($\Delta_\gamma,\Delta_Z,x_\gamma$) parameter space in which there
can be discovery as well as confusion between $Z^\prime$ and (in this case)
the $x_\gamma$-AGC model. The planar surface delimited by the solid ellipse
is determined by the projection of such three-dimensional surface,
hence of the corresponding confusion region, onto the plane
($\Delta_\gamma,\Delta_Z$). Any determination of $\Delta_\gamma$ and $\Delta_Z$
in the planar domain exterior to the ellipse would allow both $Z^\prime$
discovery and identification against the $x_\gamma$-AGC model. Similar
to the case of discovery, also in the case of $Z^\prime$ identification the
bounds on $\Delta_\gamma$ and $\Delta_Z$ could be translated into limits
on the $Z$-$Z'$  mixing angle $\phi$ and mass $M_2$ for any specific $Z'$ model.

\begin{figure}[!htb]
\includegraphics[scale=0.67]{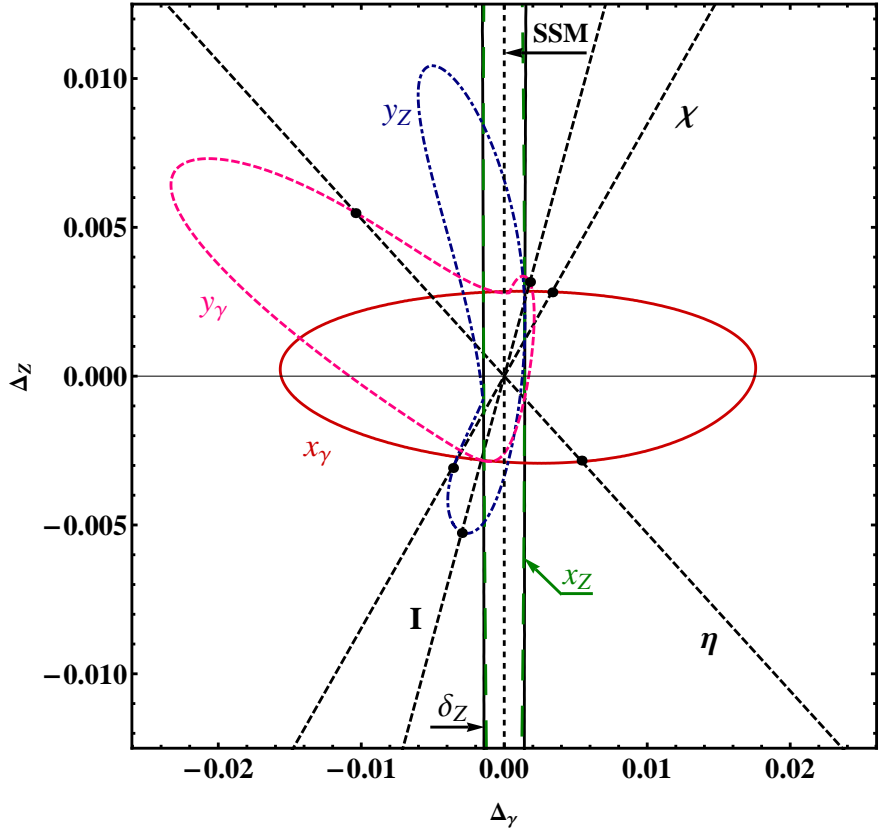}\vspace{-3mm}
\caption{The closed contours indicate regions of $(\Delta_\gamma,\Delta_Z)$ that can be populated by variations of an AGC parameter, such as for example $x_\gamma$. 
They are thus confusion regions (95\%C.L.)
%($1 - p = 0.95$)
in the parameter plane ($\Delta_{\gamma},\Delta_Z$) for a generic $Z'$ model and AGC
models with parameters taking non-vanishing values, one at a time: $x_\gamma$,  $x_Z$,
$y_\gamma$,  $y_Z$ and $\delta_Z$. Polarized cross sections with
$P_{L}=\pm 0.8$ and $\bar{P}_{L}=\mp 0.5$ have been exploited.
 Dashed straight lines correspond to specific
$Z'$ models ($\chi$, $\psi$, $\eta$, I and LRS). Here, $\sqrt{s}=0.5~\text{TeV}$ and $\Lumint=500~\text{fb}^{-1}$.}
\label{fig11}
\end{figure}

The procedure outlined above can be repeated for all other types of models with AGC
parameters ($\delta_Z$, $x_Z$, $y_\gamma$, $y_Z$), and consequently one can evaluate
the corresponding ``confusion regions'' in the ($\Delta_{\gamma},\Delta_Z$) parameter
plane. The results of this kind of analysis are represented in Fig.~\ref{fig11}
displaying the overlap of the confusion regions (95\% C.L.) in the parameter plane
($\Delta_{\gamma},\Delta_Z$) for a generic $Z'$ vector model and AGC models with
parameters varying one at a time.

%%%%%%%%%%%%%%%%%%%%%%%%%%%%%%%%%%%%%%%%%%%%%%%%%%%%%%%%
%%%%%%%%%%%%%%%%%%%%%%%%%%%%%%%%%%%%%%%%%%%%%%%%%%%%%%%%

The resulting confusion area (obtained from the overlap of all confusion regions) turns
out to be open in the vertical direction, i.e., along the $\Delta_Z$ axis. The reason
is that the $Z'$ model defined by a particular  parameter set where
($\Delta_\gamma=0,\Delta_Z$) is indistinguishable from those originating from AGC with
the same $\delta_Z=\Delta_Z$.
Moreover, from a
comparison of the confusion region depicted in Fig.~\ref{fig11} with the corresponding
discovery reach presented in Fig.~\ref{fig8a} one can conclude that all $Z'$ models
might be discovered in the process (\ref{proc1}) with polarized beams. 
However, they may not all be {\it identified}, the reason being that  the confusion region 
shown in Fig.~\ref{fig11} is not closed, in contrast
to the reach shown in Fig.~\ref{fig8a}.

An  example relevant to the current discussion can be found in the SSM model. In fact,
from Eq.~(\ref{relation}) one can conclude that the signature space of the SSM model
in the ($\Delta_\gamma,\Delta_Z$) parameter plane extends along $\Delta_Z$. It implies
that the SSM might be discovered in the process (\ref{proc1}) but not separated
from AGC models characterized by the parameter $\Delta_Z$. More generally, those
models where the $Z'$-electron couplings satisfy the equation $v'/a'=v/a$ that, as
follows from Eq.~(\ref{deltag}), lead to $\Delta_\gamma=0$ can not be
distinguished from the AGC case in the $W^\pm$ pair production process. However, all
other $Z'$ models (apart from the considered exceptional case) described by the pair of
parameters ($\Delta_\gamma,\Delta_Z$) that are located outside
of the confusion area
shown in Fig.~\ref{fig11} can be identified. Notice that the above constraint on the
electron couplings is fulfilled for an $E_6$ model at $\beta=87^\circ$ and for an LRS
model with $\alpha_{LR}=1.36$.

%%%%%%%%%%%%%%%%%%%%%%%%%%%%%%%%%%%%%%%%%%%%%%%%%%%%%%%%%%%%%%%%
\begin{figure}[!htb]
\includegraphics[scale=0.65]{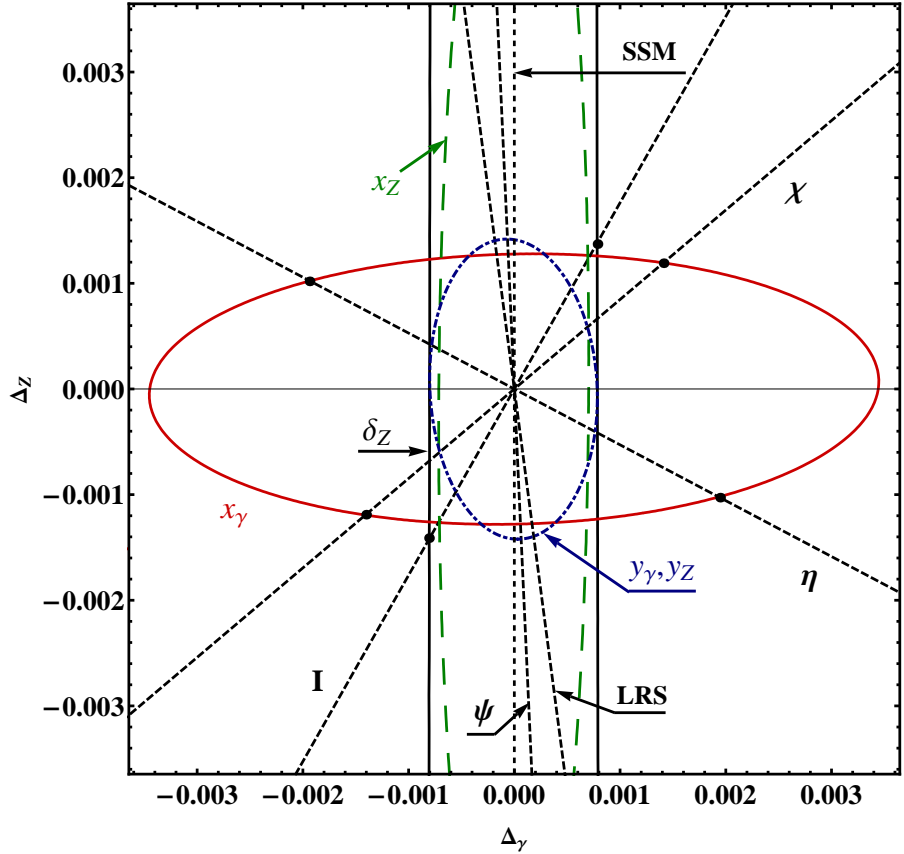}\vspace{-3mm}
\caption{Same as in Fig.~\ref{fig11} but obtained from combined
analysis of the process (\ref{proc1}) with polarized initial beams
and polarized $W^\pm$ final states. The $x_Z$ contour closes at $\Delta_Z\simeq\pm0.006$.
%$e^+(\bar{P})e^-(P)\to
%W_L^+W_L^-;\, W_L^+W_T^- +W_T^+W_L^-$.
}\label{fig11a}
\end{figure}
%%%%%%%%%%%%%%%%%%%%%%%%%%%%%%%%%%%%%%%%%%%%%%%%%%%%%%%%%%%%%%%%

The results of a further potential extension of the present analysis are
presented in Fig.~\ref{fig11a} where the feasibility of measuring
polarized $W^\pm$ states in the process (\ref{proc1}) is assumed.
This assumption is based on the experience gained at LEP2 on
measurements of $W$ polarisation \cite{LEP2Wpolar}. The relevant
theoretical framework for measurement of $W^\pm$ polarisation was
described in \cite{gounaris,Gounaris:1992kp}. The method exploited for the measurement
of $W$ polarisation is based on the spin density matrix elements
that allow to obtain the differential cross sections for polarised
$W$ bosons. Information on spin density matrix elements as
functions of the $W^-$ production angle with respect to the
electron beam direction was extracted from the decay angles of the
charged lepton in the $W^-$ ($W^+$) rest frame.

\subsection{Model dependent analysis}

\begin{figure}[htb]
\begin{center}
\includegraphics[scale=0.43]{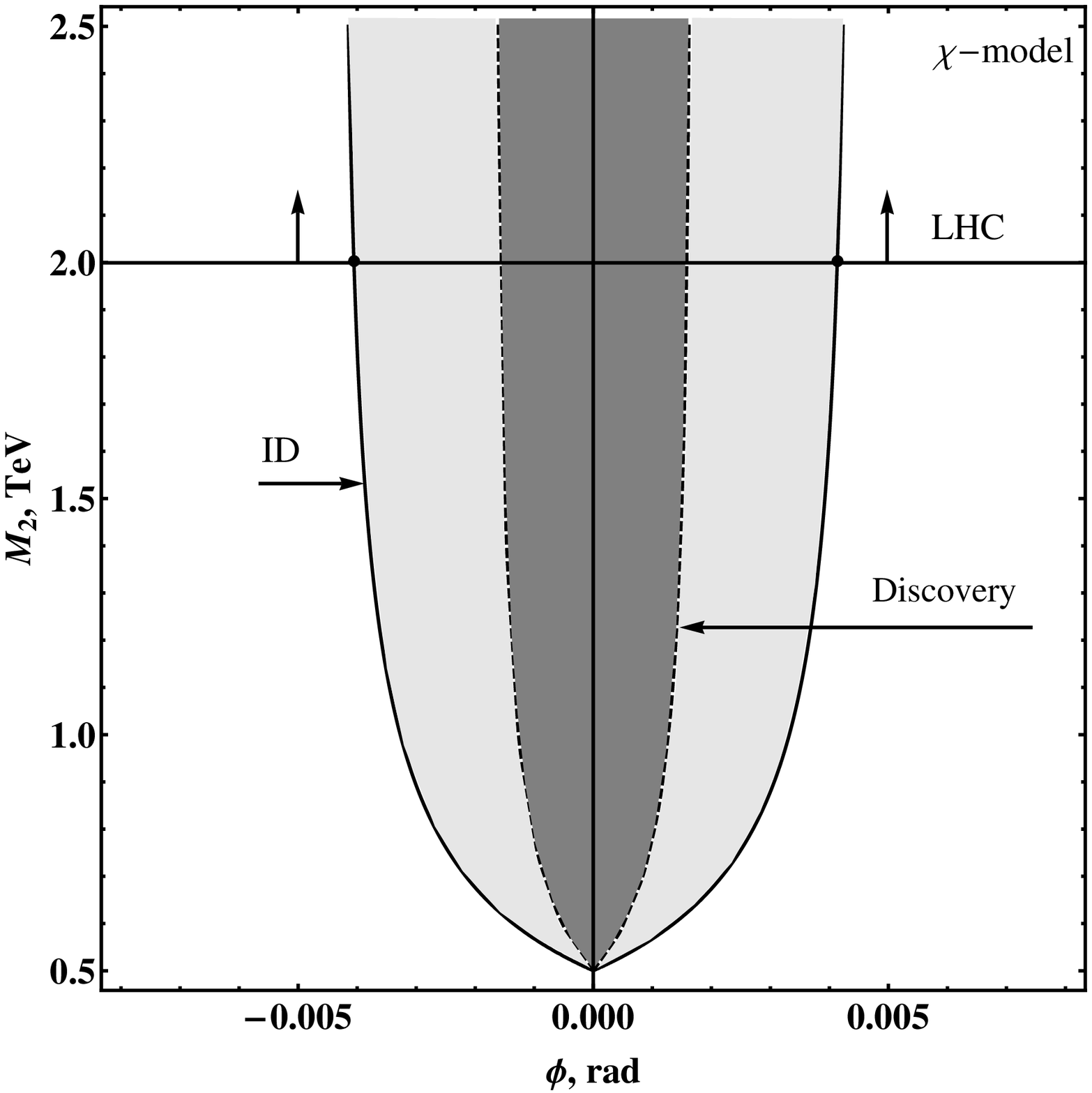}
\includegraphics[scale=0.43]{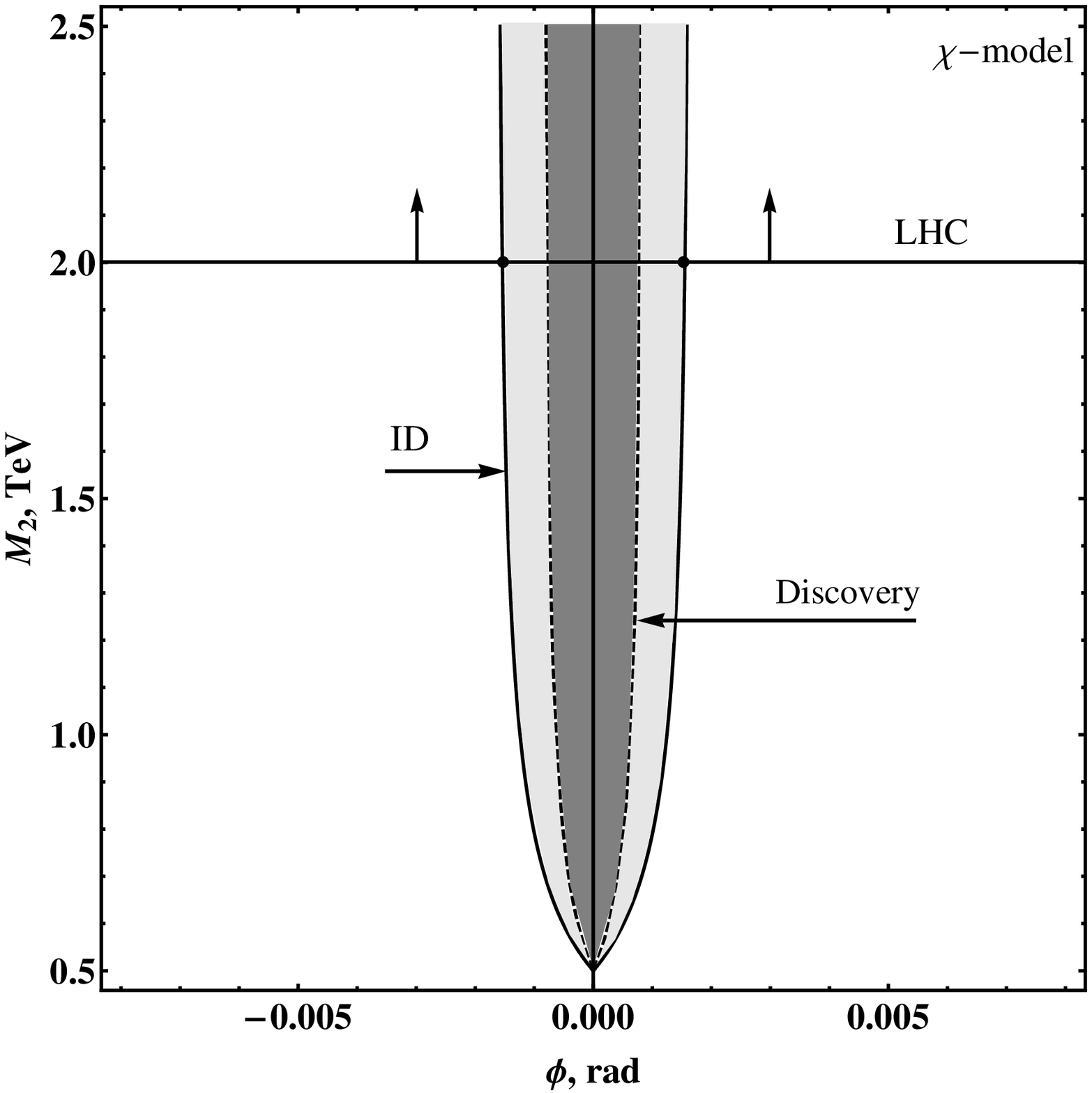}
\vspace{-4mm}
\caption{Left: Discovery (dashed line) and identification
(solid line) reach for the $\chi$ model in the $\left(\phi, M_{2}\right)$
plane obtained from polarized initial $e^+$ and $e^-$ beams with
($P_{L}=\pm 0.8,\;\bar{P}_{L}=\mp 0.5$) and unpolarized final
$W^\pm$ states.
Right: The same with polarized final $W^\pm$ states.
Here, $\sqrt{s}=0.5~\text{TeV}$ and $\Lumint=500~\text{fb}^{-1}$.
The horizontal line with vertical arrows, here and in the next figures, approximately indicates the range of $M_2$ currently 
allowed by LHC.}
\label{fig12}
\end{center}
\end{figure}

\begin{figure}[htb]
\begin{center}
\includegraphics[scale=0.43]{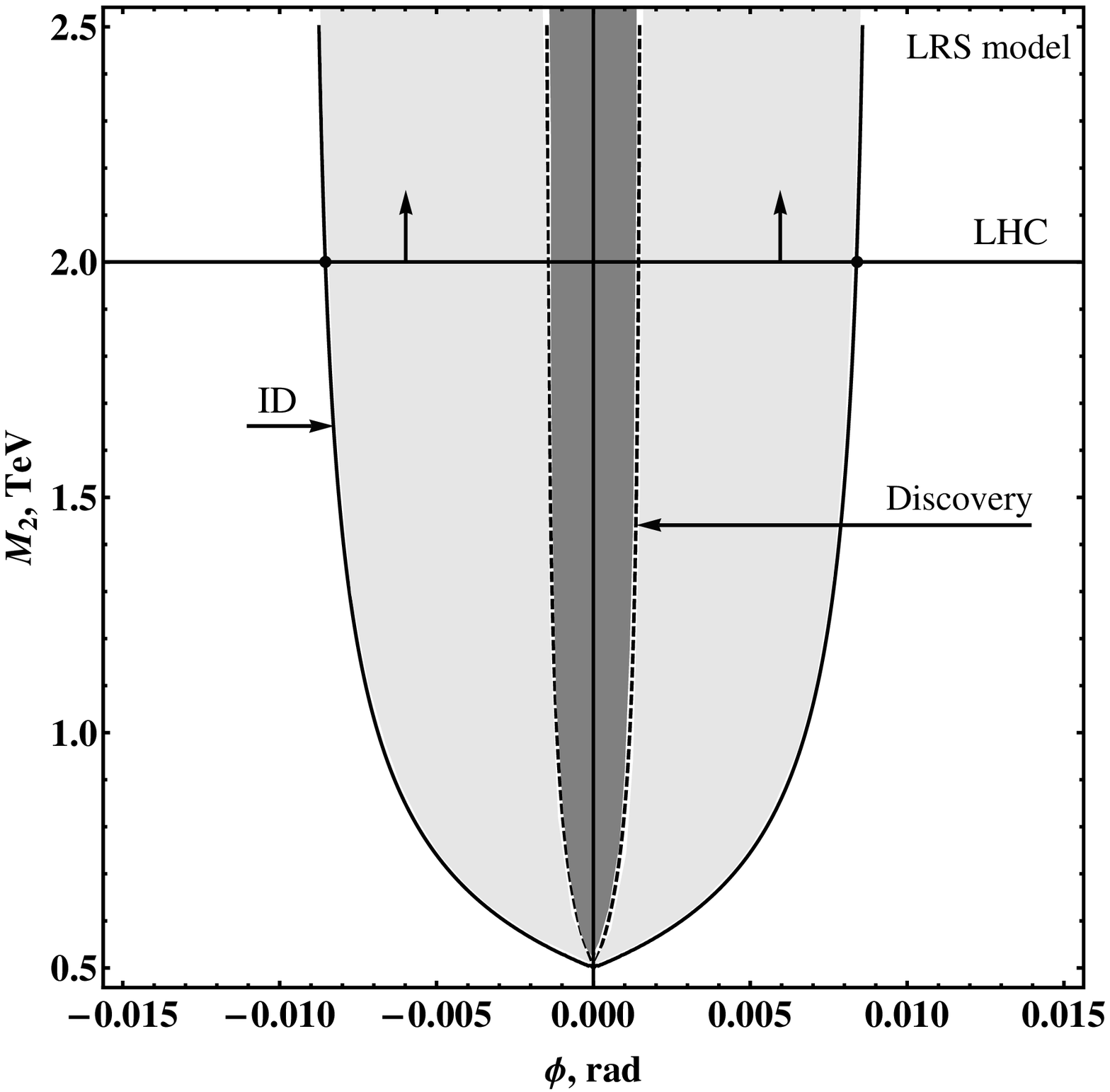}
\includegraphics[scale=0.43]{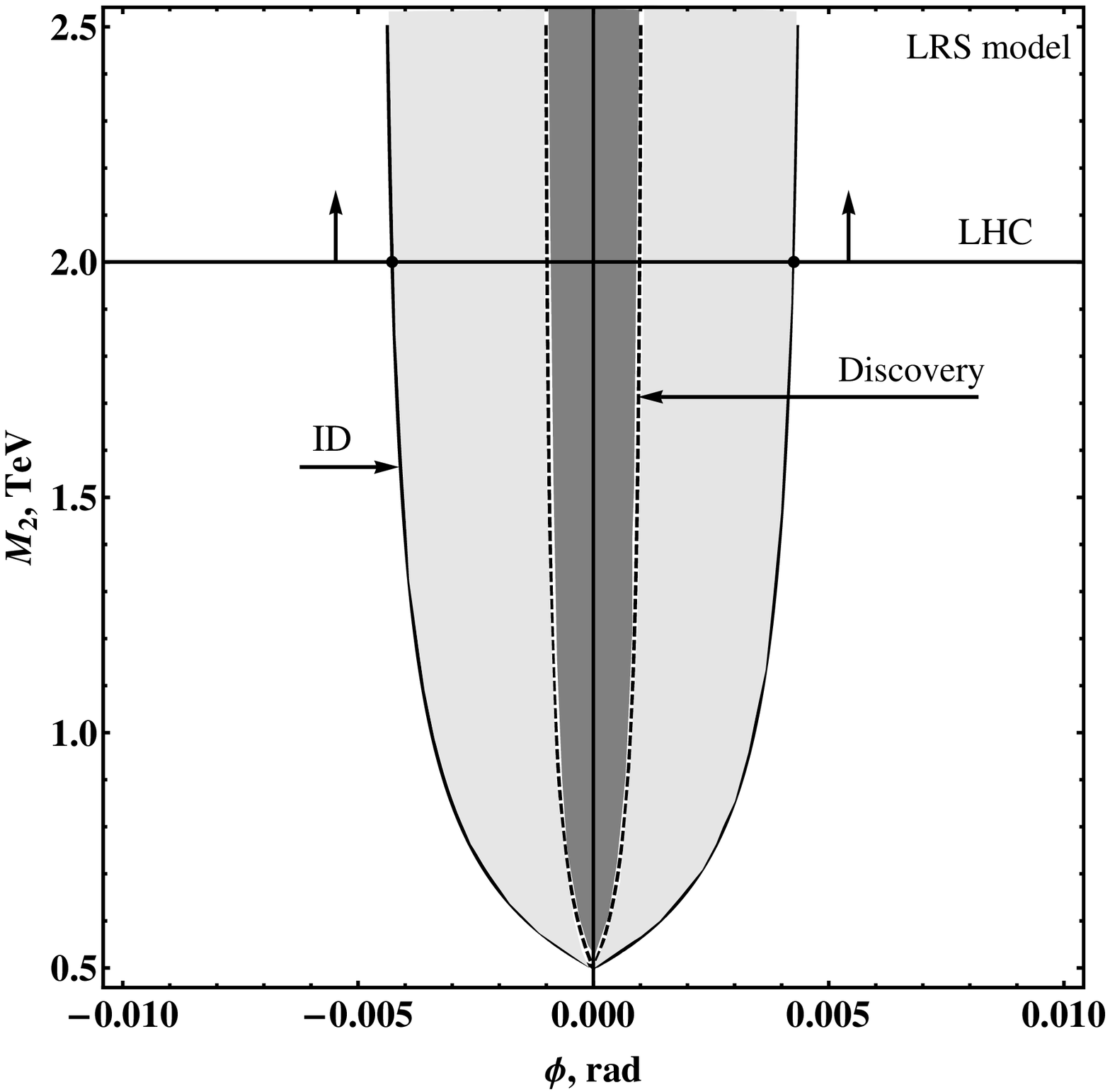}
\vspace{-4mm}\caption{Same as in Fig.\ref{fig12} but for the LRS model. }
\label{fig13}
\end{center}
\end{figure}

As mentioned above, the ranges of $\Delta_\gamma$ and $\Delta_Z$ allowed to the
specific models in Figs.~\ref{fig11} and \ref{fig11a} can be translated into discovery
and identification reaches  on the mixing angle $\phi$ and the heavier gauge boson
mass $M_{2}$, using Eqs.~(\ref{deltag})--(\ref{delta_z}). The resulting allowed
regions, discovery and identification (at the 95\% CL) in the ($\phi,M_{2}$) plane is
limited in this case by the thick dashed and solid lines, respectively, in
Figs.~\ref{fig12}-- \ref{fig13} for some specific $E_6$ models. These limits are
obtained from the polarized differential distributions of $W$ with collider energy
$\sqrt{s}=0.5~\text{TeV}$ and integrated luminosity $\Lumint=500~\text{fb}^{-1}$.
Also, an indicative typical lower bound on $M_{2}$ from direct searches at the LHC with $\sqrt{s}=7~\text{TeV}$ 
\cite{Chatrchyan:2012it,atlas-dilepton} is
reported in these figures as horizontal straight lines. The 
vertical arrows then indicate the range of available 
$Z^\prime$ mass values according to LHC limits.

%%%%%%%%%%%%%%%%%%%%%%%%%%%%%%%%%%%%%%%%%%%%%%%%%%%%%%%%%%%%%%%%
%%%%%%%%%%%%%%%%%%%%%%%%%%%%%%%%%%%%%%%%%%%%%%%%%%%%%%%%%%%%%%%%

%%%%%%%%%%%%%%%%%%%%%%%%%%%%%%%%%%%%%%%%%%%%%%%%%%%%%%%%%%%%%%%%
%%%%%%%%%%%%%%%%%%%%%%%%%%%%%%%%%%%%%%%%%%%%%%%%%%%%%%%%%%%%%%%%

%%%%%%%%%%%%%%%%%%%%%%%%%%%%%%%%%%%%%%%%%%%%%%%%%%%%%%%%%%%%%%%%
\begin{figure}[htb]
\begin{center}
\includegraphics[scale=0.43]{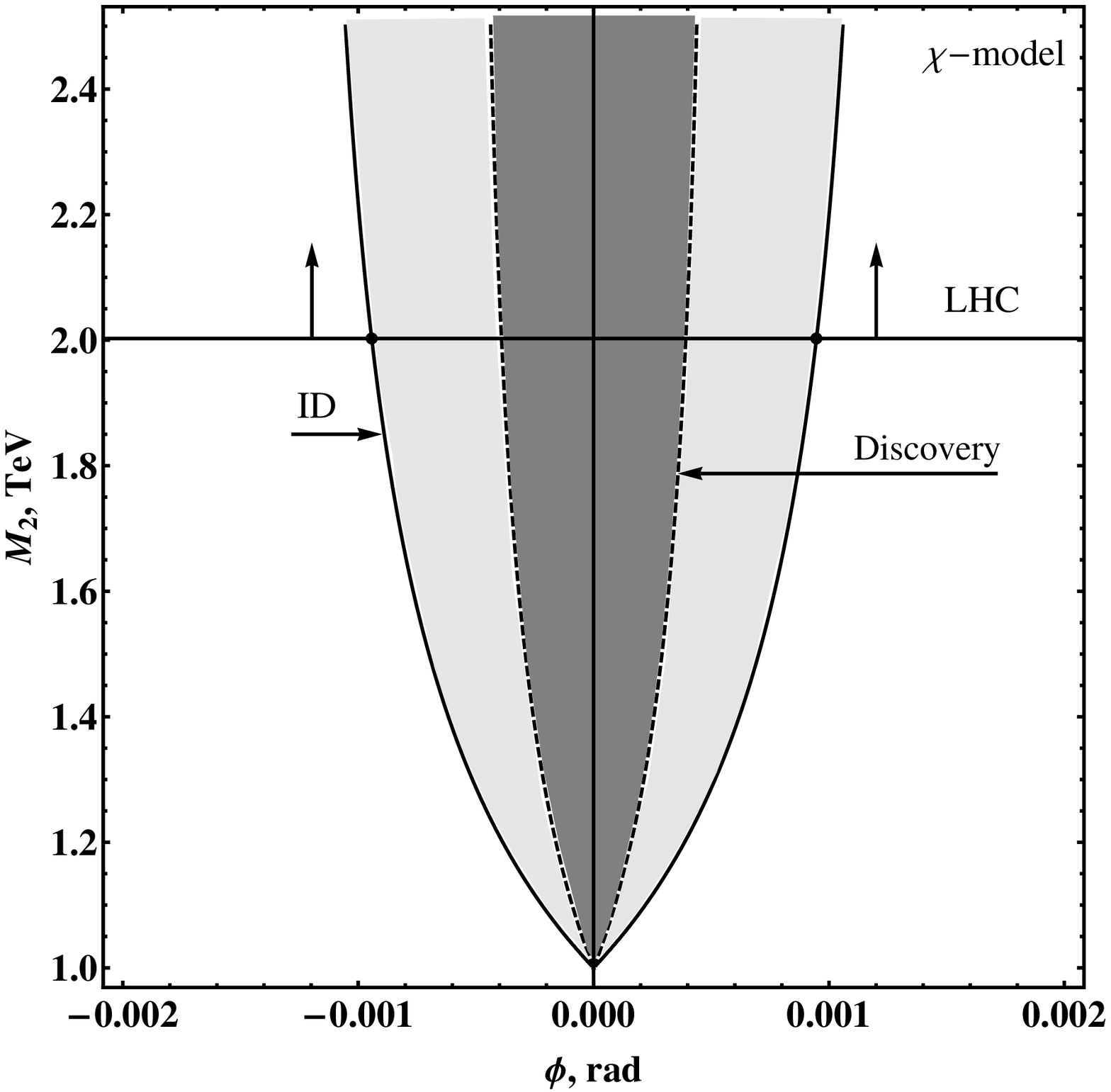}
\includegraphics[scale=0.43]{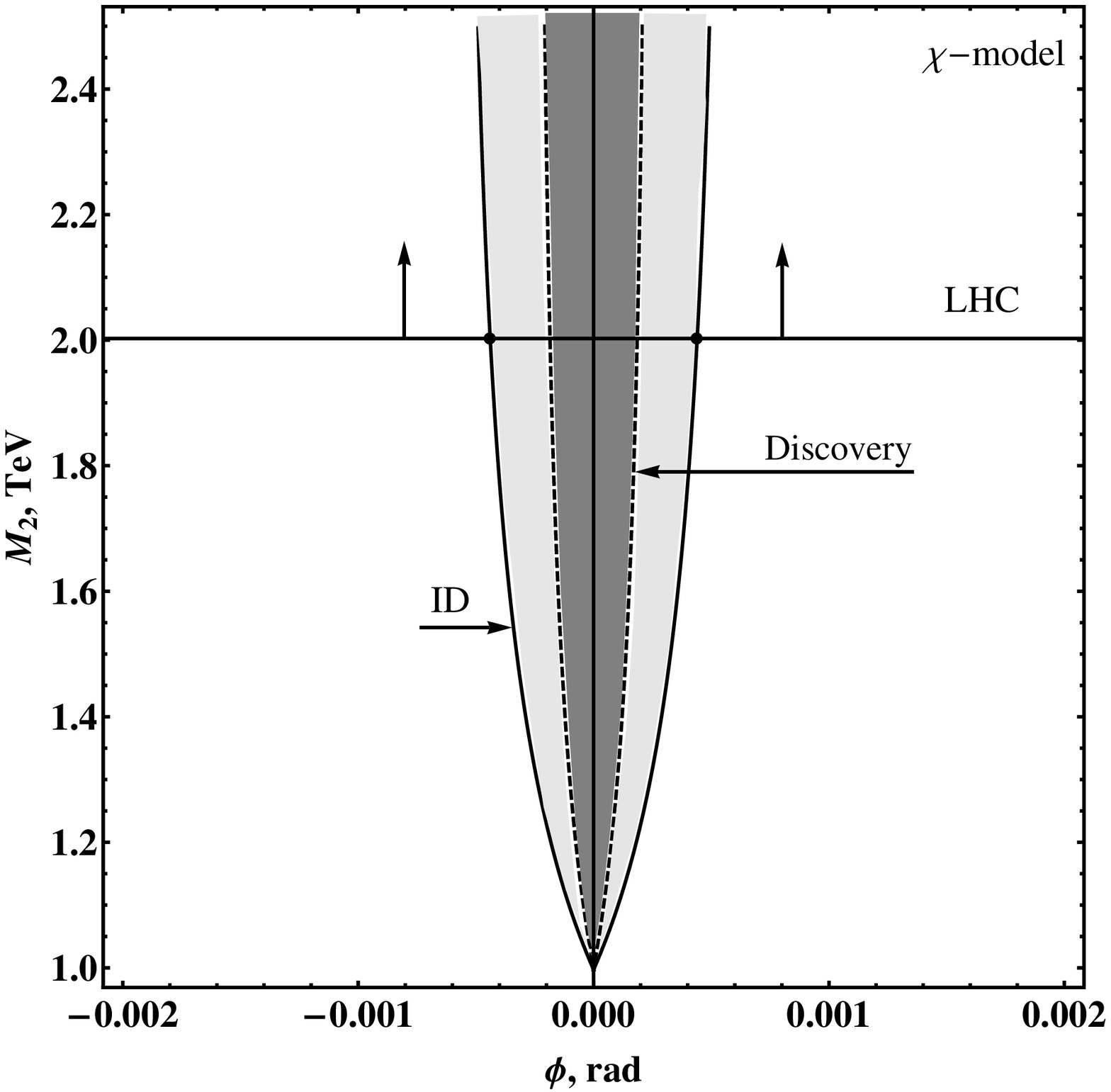}
\vspace{-4mm} \caption{Left: Discovery (dashed line) and identification (solid line)
reach for the $\chi$ model in the $\left(\phi, M_{2}\right)$ plane obtained from
polarized initial $e^+$ and $e^-$ beams with ($P_{L}=\pm 0.8,\;\bar{P}_{L}=\mp 0.5$)
and unpolarized final $W^\pm$ states. Right: The same with polarized final $W^\pm$
states. Here, $\sqrt{s}=1~\text{TeV}$ and $\Lumint=1~\text{ab}^{-1}$.} \label{fig14}
\end{center}
\end{figure}
%%%%%%%%%%%%%%%%%%%%%%%%%%%%%%%%%%%%%%%%%%%%%%%%%%%%%%%%%%%%%%%%

%%%%%%%%%%%%%%%%%%%%%%%%%%%%%%%%%%%%%%%%%%%%%%%%%%%%%%%%%%%%%%%%
\begin{figure}[htb]
\begin{center}
\includegraphics[scale=0.43]{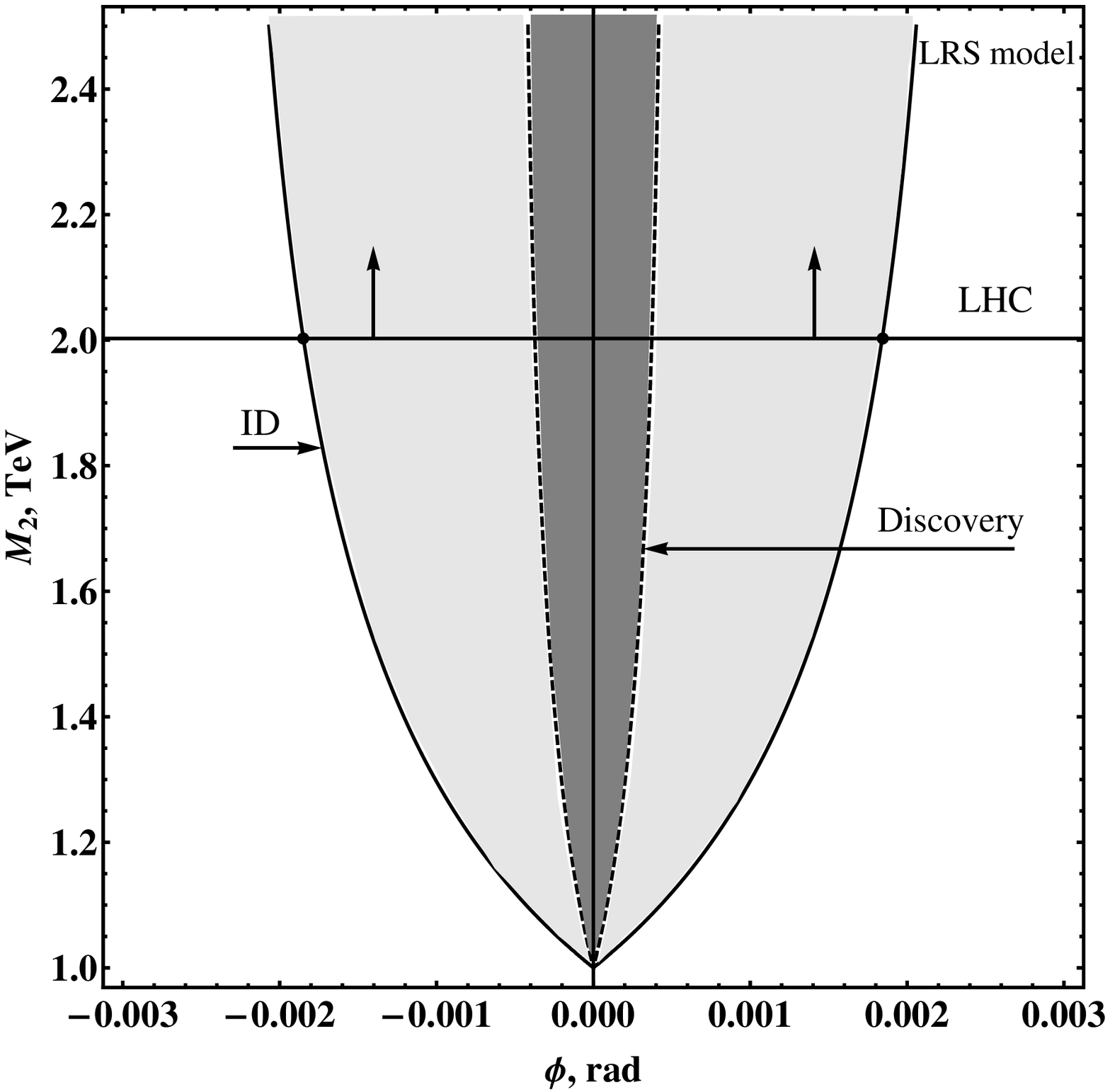}
\includegraphics[scale=0.43]{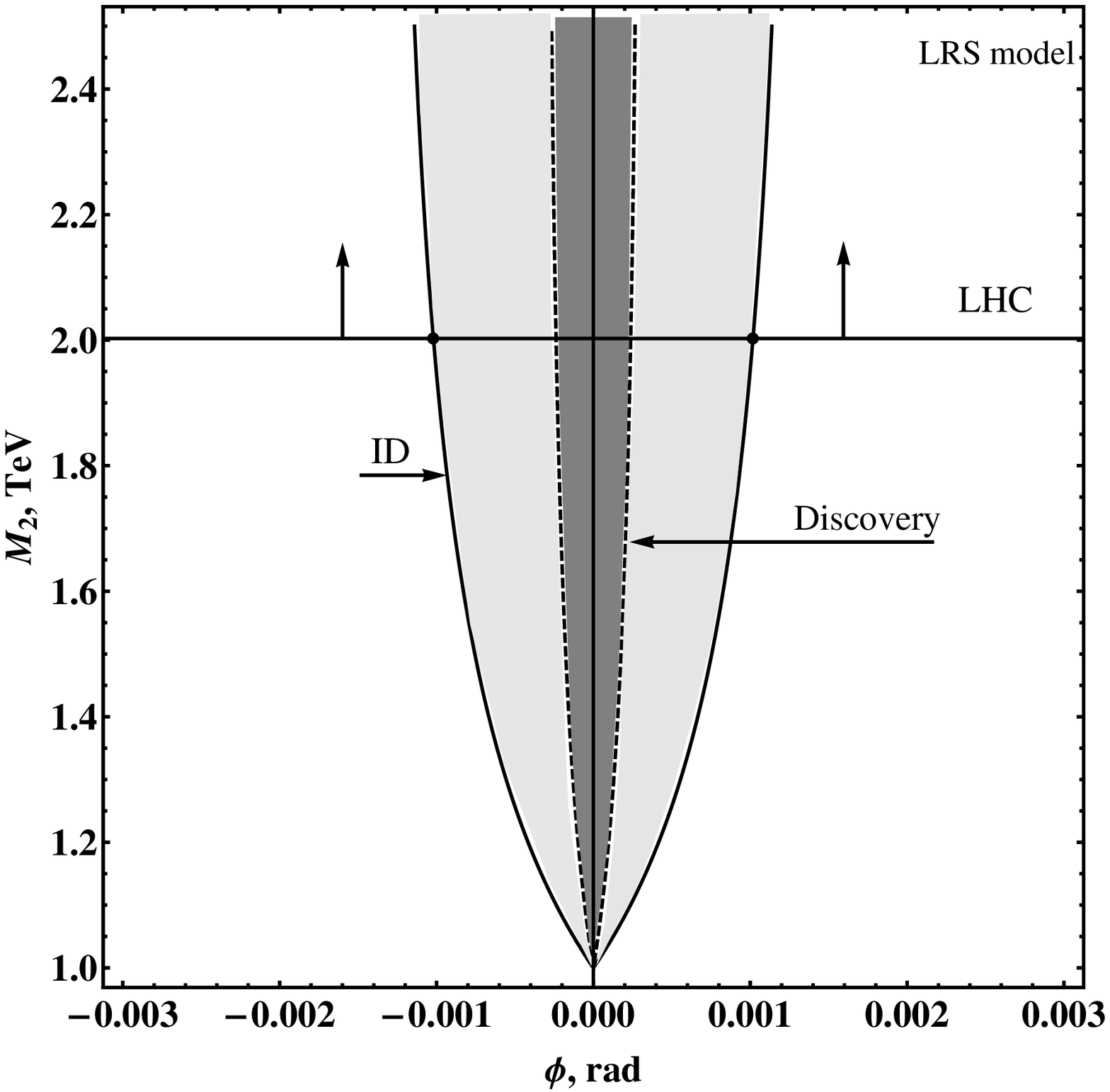}
\vspace{-4mm}\caption{Same as in Fig.\ref{fig14} but for the LRS model.} \label{fig15}
\end{center}
\end{figure}
%%%%%%%%%%%%%%%%%%%%%%%%%%%%%%%%%%%%%%%%%%%%%%%%%%%%%%%%%%%%%%%%

%%%%%%%%%%%%%%%%%%%%%%%%%%%%%%%%%%%%%%%%%%%%%%%%%%%%%%%%%%%%%%%%
\begin{table}[h!]
\caption{Discovery and identification reach on the $Z$-$Z'$ mixing angle $\phi$ for
$Z'$ models with $M_{2}=2~\text{TeV}$ obtained from the polarized differential cross
section with
 ($P_{L}=\pm 0.8,\;\bar{P}_{L}=\mp 0.5$) and unpolarized final
states for the case $\sqrt{s}=0.5~\text{TeV}$ and $\Lumint=500~\text{fb}^{-1}$. 
The corresponding limits for polarized $W$s are given in
parenthesis. } \vspace{1mm}
\begin{center}
\begin{tabular}{|c|c|c|c|c|c|c|}
\hline $Z^\prime$ model & $\chi$ & $\psi$ & $\eta$ & I & LRS & SSM \\
\hline $\phi^\text{DIS}, 10^{-3}$
& $\pm 1.5 (0.8)$ & $\pm 2.3 (1.4)$ & $\pm 1.6 (1.3)$ & $\pm 2.0 (0.8)$ & $\pm 1.4 (1.0)$ & $\pm 1.2 (0.7)$ \\
\hline $\phi^\text{ID}, 10^{-3}$
& $\pm 3.8 (1.5)$ & $\pm 36.8 (18.5)$ & $\pm 17.4 (3.2)$ & $\pm 4.3 (1.2)$ & $\pm 8.1 (4.2)$ & -- \\
\hline
\end{tabular}
\end{center}
\label{tab3}
\end{table}
%%%%%%%%%%%%%%%%%%%%%%%%%%%%%%%%%%%%%%%%%%%%%%%%%%%%%%%%%%%%

Figures~\ref{fig12} and \ref{fig13} show that the process $e^+e^-\to W^+W^-$ at
$0.5~\text{TeV}$ has a potential sensitivity to the mixing angle $\phi$ of the order
of $10^{-4}$--$10^{-3}$ or even less, depending on the mass $M_{2}$. This sensitivity would increse for 
the c.m.\ energy $\sqrt{s}$ approaching $M_2$ because the contribution of the $Z_2$ exchange diagram in 
Fig.~\ref{fig2} would be enhanced. However, $Z'$ bosons relevant to the extended
models under study with mass below $\sim 2.0-2.3~\text{TeV}$ are already excluded by LHC data, and the ILC c.m.\ energies  considered here are therefore quite far from the admissible 
$M_2$. Conversely, for masses $M_{2}$ much larger than $\sqrt{s}$ such that the
$Z_2$ exchange contribution $\vert\chi_2/\chi\vert$ is much less than unity, the
limiting contour is mostly determined by the modification (\ref{v1}) of the $Z$ couplings to electrons. 
The discovery and identification reaches on $\phi$ at
$M_2=2~\text{TeV}$ are summarized in Table~\ref{tab3}.

For the ILC with higher energy and luminosity, $\sqrt{s}=1~\text{TeV}$
and $\Lumint=1$ ab$^{-1}$, one expects further improvement of the
discovery and identification reach on the $Z$-$Z'$ mixing angle and
$M_2$ (see Figures~\ref{fig14}, \ref{fig15} and Table~\ref{tab4}).

\begin{table}[h!]
\caption{ Same as in Table~\ref{tab3} but for ILC with
 $\sqrt{s}=1~\text{TeV}$ and $\Lumint=1~\text{ab}^{-1}$.} \vspace{1mm}
\begin{center}
\begin{tabular}{|c|c|c|c|c|c|c|}
\hline $Z^\prime$ model & $\chi$ & $\psi$ & $\eta$ & I & LRS & SSM \\
\hline $\phi^\text{DIS}, 10^{-4}$
& $\pm 3.8 (1.8)$ & $\pm 5.8 (3.4)$ & $\pm 4.6 (3.2)$ & $\pm 4.4 (1.9)$ & $\pm 3.7 (2.4)$ & $\pm 3.1 (1.7)$ \\
\hline $\phi^\text{ID}, 10^{-4}$
& $\pm 9.0 (4.2)$ & $\pm 94 (45)$ & $\pm 24 (9.5)$ & $\pm 6.1 (2.8)$ & $\pm 18 (10)$ & -- \\
\hline
\end{tabular}
\end{center}
\label{tab4}
\end{table}

As already mentioned, the horizontal lines in Figs.~\ref{fig12}--\ref{fig15} denote the current LHC lower limits on  $M_2$, therefore only the upper parts, as indicated  by the vertical arrows,  will be available for discovery and identification of a $Z'$ {\it via} indirect manifestations at the ILC with the considered values for the c.m.\ energy of 0.5 and 1~TeV. Since
those limits are so much higher than $\sqrt{s}$, the corrections from finite $Z'$
widths,  assumed in the range $\Gamma_{Z'}=(0.01-0.10) M_{Z'}$ \cite{Langacker:2008yv}, are found to
be  numerically negligible in the ``working" regions indicated in those
figures  by the horizontal lines and vertical arrows.
Tables~\ref{tab3} and \ref{tab4} demonstrate that  ILC~(0.5~TeV) and ILC~(1 TeV)
allow to improve current bounds on $Z$--$Z'$ mixing for most of the $Z'$ models, and also
differentiating $Z'$ from AGC is feasible.

%%%%%%%%%%%%%%%%%%%%%%%%%%%%%%%%%%%%%%%%%%%%%%%%%%%%%%%%%%%%%%%%
\subsection{Low-energy option}
%%%%%%%%%%%%%%%%%%%%%%%%%%%%%%%%%%%%%%%%%%%%%%%%%%%%%%%%%%%%%%%%

Currently, physics at the ILC in a low-energy option
is extensively studied and discussed, as it in this mode might act as a ``Higgs factory''.
The results for discovery and identification reach on $Z$-$Z'$ mixing
and mass $M_2$ obtained from the ILC with  $\sqrt{s} =0.25$ TeV
and 0.35 TeV are summarized in
Tables~\ref{tab5} and \ref{tab6}.

\begin{table}[h!]
\caption{ Same as in Table~\ref{tab3} but for the ILC with
  $\sqrt{s}=0.25~\text{TeV and}$ $\Lumint=100$ fb$^{-1}$.}
\begin{center}
\begin{tabular}{|c|c|c|c|c|c|c|}
\hline $Z^\prime$ model & $\chi$ & $\psi$ & $\eta$ & I & LRS & SSM \\
\hline $\phi^\text{DIS}, 10^{-3}$
& $\pm 5.1 (3.8)$ & $\pm 8.4 (7.0)$ & $\pm 6.8 (6.7)$ & $\pm 5.7 (3.9)$ & $\pm 5.4 (4.9)$ & $\pm 4.4 (3.6)$ \\
\hline $\phi^\text{ID}, 10^{-3}$
& $\pm 14 (6.8)$ & $\pm 109 (86)$ & $\pm 29 (14)$ & $\pm 7.8 (5.9)$ & $\pm 45 (21)$ & -- \\
\hline
\end{tabular}
\end{center}
\label{tab5}
\end{table}

\begin{table}[h!]
\caption{Same as in Table~\ref{tab5} but for the ILC with
 $\sqrt{s}=0.35~\text{TeV}$, and two values of integrated luminosity.}
\begin{center}
\begin{tabular}{|c|c|c|c|c|c|c|c|}
\hline 
\multicolumn{2}{|c|}{$Z^\prime$ model} & $\chi$ & $\psi$ & $\eta$ & I & LRS & SSM \\
\hline 
\multirow{2}{*}{$100~\text{fb}^{-1}$}
&$\phi^\text{DIS}, 10^{-3}$
& $\pm 3.7 (2.4)$ & $\pm 6.0 (4.5)$ & $\pm 4.9 (4.3)$ & $\pm 4.1 (2.5)$ & $\pm 3.9 (3.1)$ & $\pm 3.2 (2.3)$ \\
& $\phi^\text{ID}, 10^{-3}$
& $\pm 8.4 (4.6)$ & $\pm 77 (61)$ & $\pm 27 (9.4)$ & $\pm 13.5 (3.8)$ & $\pm 19 (14)$ & -- \\
\hline
\hline 
\multirow{2}{*}{$500~\text{fb}^{-1}$}
& $\phi^\text{DIS}, 10^{-3}$
& $\pm 2.3 (1.3)$ & $\pm 3.4 (2.3)$ & $\pm 2.5 (2.1)$ & $\pm 3.1 (1.4)$ & $\pm 2.1 (1.6)$ & $\pm 1.8 (1.2)$ \\
& $\phi^\text{ID}, 10^{-3}$
& $\pm 5.9 (2.4)$ & $\pm 54 (30)$ & $\pm 15 (4.7)$ & $\pm 4.0 (1.9)$ & $\pm 16 (6.8)$ & -- \\
\hline
\end{tabular}
\end{center}
\label{tab6}
\end{table}

The comparison of these constraints with those obtained from electroweak precision
data derived mostly from on-$Z$-resonance experiments at LEP1 and SLC
\cite{Erler:2009jh} shows that the ILC~(0.25~TeV) and ILC~(0.35~TeV) allow to obtain
bounds on $Z$-$Z^\prime$ mixing at the same level as those of current experimental limits, thereby
providing complementary bounds on $Z'$s. 

Increasing the luminosity at fixed energy, asymptotically allows for an increase of the sensitivity $\propto1/\sqrt{\Lumint}$. In the example shown in Table~\ref{tab6}, this behavior is not quite reached, due to the impact of systematic uncertainties.

\section{Concluding remarks}

We have discussed  the foreseeable sensitivity to $Z'$s in $W^\pm$-pair
production cross sections at the ILC, especially as regards the
potential of distinguishing  observable effects of a $Z'$ from
analogous ones due to competitor models with Anomalous Gauge Couplings that can lead to
the same or similar new physics experimental signatures.
The discovery and identification reaches on $E_6$ and LRS models have been determined in the parameter plane spanned by the $Z$-$Z^\prime$ mixing angle $\phi$, and $Z^\prime$ mass, $M_2$.

 We have shown that the sensitivity of the ILC for probing the $Z$-$Z'$ mixing and its
capability to distinguish these two new physics scenarios is
substantially enhanced when the polarization of the initial beams
(and also, possibly, the produced $W^\pm$ bosons)  are considered.

\section*{Acknowledgements}
\par\noindent
It is a pleasure to thank S. Dittmaier for valuable comments on the importance of the radiative corrections.
 This research has been partially supported by the Abdus
Salam ICTP under the TRIL and STEP Programmes and the Belarusian Republican Foundation
for Fundamental Research. The work of AAP  has been partially supported by the SFB 676
Programme of the Department of Physics, University of Hamburg. The work of PO has been
supported by the Research Council of Norway.

%%%%%%%%%%%%%%%%%%%%%%%%%%%%%%%%%%%%%%%%%%%%%%%%%%%%%%%%%%%%%%%%
\section*{Appendix~A. Helicity amplitudes}
\setcounter{equation}{0}
\renewcommand{\thesection}{A}

In this appendix, we collect the helicity amplitudes for the different initial ($e^+e^-$) and final-state ($W^+W^-$) polarizations. In Table~\ref{tab_amplit} we quote the amplitudes for the case of Anomalous Gauge Couplings \cite{gounaris,Gounaris:1992kp}, whereas in Table~\ref{tab:tab1b} we give the corresponding results for the case of a $Z^\prime$.

%__________________________________________________
%\hskip -1 cm
\begin{table}[h]
\centering \caption{Helicity amplitudes for $e^+e^-\to W^+W^-$ in the presence of AGC  \cite{gounaris,Gounaris:1992kp}. 
To obtain the amplitude $F_{\lambda\tau\tau'}(s,\cos\theta)$ for
definite helicity $\lambda=\pm 1/2$ and definite spin orientations $\tau(W^-)$ and $\tau'(W^+)$ of the
$W^\pm$, the elements in the corresponding column have to be multiplied by the common
factor on top of the column. Subsequently, the elements in a specific column have to
be multiplied by the corresponding elements in the first column and the sum over all
elements is to be taken. In the last column, the amplitude for the case of $\tau=\pm
1$, $\tau^{\prime}=0$ is obtained by replacing $\tau^{\prime}$ by $-\tau$ in the
elements of this last column.} \vspace{3mm}
\begin{tabular}{|c|c|c|}
\hline
\multirow{2}{*}{$e^+_{-\lambda} e^-_\lambda\to W^+_LW^-_L$}  &
$ \tau=\tau^\prime=0$&$$\\
$$&$-\frac{e^2 s\lambda}{2}\sin\theta$ & $$\\ \hline
$\frac{2\lambda-1}{4\hskip 2pt t\hskip 2pt s^2_W}$ &
$\frac{s}{2M_W^2}[\cos\theta-\beta_W (1+\frac{2M_W^2}{s})]$ &\\
\hline $-\frac{2}{s}+
\frac{2(\cot\theta_W+\delta_Z)}{s-M_Z^2}(v-2a\lambda)$ &
$-\beta_W(1+\frac{s}{2M^2_W})$ &\\ \hline $ -\frac{x_\gamma}{s}+
\frac{x_Z}{s-M_Z^2}(v-2a\lambda)$ & $-\beta_W\frac{s}{M^2_W}$ &\\
\hline \hline
\multirow{2}{*}{$e^+_{-\lambda} e^-_\lambda\to W^+_TW^-_T$} &
$\tau=\tau^\prime=
\pm 1$ & $\tau=-\tau^\prime=\pm 1$ \\
& $-\frac{e^2 s\lambda}{2}\sin\theta$
& $-\frac{e^2 s\lambda}{2}\sin\theta$\\
\hline $\frac{2\lambda-1}{4\hskip 2pt t\hskip 2pt s^2_W}$ &
$\cos\theta-\beta_W $ & $-\cos\theta-2\tau\lambda $\\
\hline
$-\frac{2}{s}+\frac{2(\cot\theta_W+\delta_Z)}
{s-M_Z^2}(v-2a\lambda)$ & $-\beta_W $ & $0$ \\ \hline
$-\frac{y_\gamma}{s}+ \frac{y_Z}{s-M_Z^2}(v-2a\lambda)$ &
$-\beta_W\frac{s}{M^2_W}$ & $0$ \\ \hline \hline
\multirow{2}{*}{$e^+_{-\lambda}e^-_\lambda\to W^+_TW^-_L$} &
$\tau=0$, $\tau^\prime=\pm 1$ &
$\tau=\pm 1$, $\tau^\prime=0$ \\
& $-\frac{e^2
s\lambda}{2\sqrt{2}}(\tau^\prime\cos\theta-2\lambda)$
& $\frac{e^2 s\lambda}{2\sqrt{2}}(\tau\cos\theta+2\lambda)$\\
\hline
\multirow{2}{*}{$\frac{2\lambda-1}{4\hskip 2pt t\hskip 2pt s^2_W}$} &
$\frac{\sqrt{s}}{2M_W}[\cos\theta(1+\beta_W^2)-2\beta_W] $ &
$\frac{\sqrt{s}}{2M_W}[\cos\theta(1+\beta_W^2)-2\beta_W]$\\
& $-\frac{2M_W}{\sqrt{s}}\frac{\tau^\prime\sin^2\theta}
{\tau^\prime\cos\theta-2\lambda}$ &
$-\frac{2M_W}{\sqrt{s}}\frac{\tau\sin^2\theta}
{\tau\cos\theta+2\lambda}$ \\
\hline
$-\frac{2}{s}+\frac{2(\cot\theta_W+\delta_Z)}
{s-M_Z^2}(v-2a\lambda)$ & $-\beta_W\frac{\sqrt{s}}{M_W}$ &
$-\beta_W\frac{\sqrt s}{M_W}$ \\
\hline $-\frac{x_\gamma+y_\gamma}{s}+
\frac{x_Z+y_Z}{s-M_Z^2}(v-2a\lambda)$ & $-\beta_W\frac{\sqrt
s}{M_W}$ &
$-\beta_W\frac{\sqrt s}{M_W}$ \\
\hline
\end{tabular}
\label{tab_amplit}
\end{table}
%------------------------------------------------------------------------------

Note that the quantity $\delta_Z$ appearing in Table~\ref{tab_amplit} is different from, but plays a  role similar to that of $\Delta_Z$ entering in the parametrization of $Z^\prime$ effects. Furthermore, in analogy with the $\Delta_\gamma$ which enters the description of $Z^\prime$ effects, one could imagine a factor $(1+\delta_\gamma)$ multiplying the photon-exchange amplitudes in Table~\ref{tab_amplit}. Such a term could be induced by dimension-8 operators, but $\delta_\gamma$ would have to vanish as $s\to0$, due to gauge invariance.

%___________________________________________________________________________
%\hskip -1 cm
\begin{table}[htb!]
\centering
\caption{Helicity amplitudes for  $e^+e^-\to\gamma,Z_1,Z_2\to
W^+W^-$. } \label{tab:tab1b}
\begin{tabular}{|c|c|c|}
\hline
$$ $e^+_{-\lambda} e^-_\lambda\to W^+_LW^-_L$ & $\tau=\tau^\prime=0$ & \\
& $-\frac{e^2 s\lambda}{2}\sin\theta$ & \\
\hline $\frac{2\lambda-1}{4\hskip 2pt t\hskip 2pt s^2_W}$ &
$\frac{s}{2M_W^2}[\cos\theta-\beta_W (1+\frac{2M_W^2}{s})]$ &\\
\hline $-\frac{2}{s}+\frac{2\,
g_{WWZ_1}}{s-M_1^2+iM_1\Gamma_1}\,(v_{1}-2a_{1}\lambda) $  &
$-\beta_W(1+\frac{s}{2M^2_W})$ &\\
 $+\frac{2\,
g_{WWZ_2}}{s-M_2^2+iM_2\Gamma_2}\,(v_{2}-2a_{2}\lambda)$ && \\
$\approx -\frac{2(1+\Delta_\gamma)}{s}+
\frac{2(\cot\theta_W+\Delta_Z)}{s-M_Z^2}(v-2a\lambda)$
&& \\
\hline \hline $e^+_{-\lambda}
e^-_\lambda\to W^+_TW^-_T$ & $\tau=\tau^\prime=
\pm 1$ & $\tau=-\tau^\prime=\pm 1$ \\
& $-\frac{e^2 s\lambda}{2}\sin\theta$
& $-\frac{e^2 s\lambda}{2}\sin\theta$\\
\hline $\frac{2\lambda-1}{4\hskip 2pt t\hskip 2pt s^2_W}$ &
$\cos\theta-\beta_W $ & $-\cos\theta-2\tau\lambda $\\
\hline $-\frac{2}{s}+\frac{2\,
g_{WWZ_1}}{s-M_1^2+iM_1\Gamma_1}\,(v_{1}-2a_{1}\lambda) $ &
  $-\beta_W $ & $0$ \\
 $+\frac{2\,
g_{WWZ_2}}{s-M_2^2+iM_2\Gamma_2}\,(v_{2}-2a_{2}\lambda) $ && \\
 $\approx -\frac{2(1+\Delta_\gamma)}{s}+
\frac{2(\cot\theta_W+\Delta_Z)}{s-M_Z^2}(v-2a\lambda)$
&& \\
  \hline
\hline $e^+_{-\lambda} e^-_\lambda\to W^+_TW^-_L$ & $\tau=0$,
$\tau^\prime=\pm 1$ &
$\tau=\pm 1$, $\tau^\prime=0$ \\
& $-\frac{e^2
s\lambda}{2\sqrt{2}}(\tau^\prime\cos\theta-2\lambda)$
& $\frac{e^2 s\lambda}{2\sqrt{2}}(\tau\cos\theta+2\lambda)$\\
\hline $\frac{2\lambda-1}{4\hskip 2pt t\hskip 2pt s^2_W}$ &
$\frac{\sqrt{s}}{2M_W}[\cos\theta(1+\beta_W^2)-2\beta_W] $ &
$\frac{\sqrt{s}}{2M_W}[\cos\theta(1+\beta_W^2)-2\beta_W]$\\
& $-\frac{2M_W}{\sqrt{s}}\frac{\tau^\prime\sin^2\theta}
{\tau^\prime\cos\theta-2\lambda}$ &
$-\frac{2M_W}{\sqrt{s}}\frac{\tau\sin^2\theta}
{\tau\cos\theta+2\lambda}$ \\
\hline $-\frac{2}{s}+\frac{2\,
g_{WWZ_1}}{s-M_1^2+iM_1\Gamma_1}\,(v_{1}-2a_{1}\lambda) $  &
$-\beta_W\frac{\sqrt{s}}{M_W}$ &
$-\beta_W\frac{\sqrt s}{M_W}$ \\
 $+\frac{2\,
g_{WWZ_2}}{s-M_2^2+iM_2\Gamma_2}\,(v_{2}-2a_{2}\lambda) $
&& \\
$\approx -\frac{2(1+\Delta_\gamma)}{s}+
\frac{2(\cot\theta_W+\Delta_Z)}{s-M_Z^2}(v-2a\lambda)$
&& \\
\hline
\end{tabular}
\end{table}
%___________________________________________________________________________

\clearpage

%%%%%%%%%%%%%%%%%%%%%%%%%%%%%%%%%%%%%%%%%%%%%%%%%%%%%%%%%%%%%%%%%%%%%%%%%


\begin{thebibliography}{99}

%\cite{Langacker:2008yv}
\bibitem{Langacker:2008yv}
  P.~Langacker,
  %``The Physics of Heavy $Z^\prime$ Gauge Bosons,''
  Rev.\ Mod.\ Phys.\  {\bf 81}, 1199-1228 (2009)
  [arXiv:0801.1345 [hep-ph]].

 %\cite{Rizzo:2006nw}
\bibitem{Rizzo:2006nw}
  T.~G.~Rizzo,
  %``$Z^\prime$ phenomenology and the LHC,''
    [hep-ph/0610104].

%\cite{Leike:1996pj}
\bibitem{Leike:1996pj} 
  A.~Leike and S.~Riemann,
  %``$Z^\prime$ search in $e^{+} e^{-}$ annihilation,''
  Z.\ Phys.\ C {\bf 75}, 341 (1997)
  [hep-ph/9607306].
  %%CITATION = HEP-PH/9607306;%%

%\cite{Leike:1998wr}
\bibitem{Leike:1998wr}
  A.~Leike,
  %``The Phenomenology of extra neutral gauge bosons,''
  Phys.\ Rept.\  {\bf 317}, 143-250 (1999)
  [hep-ph/9805494].

%\cite{Riemann:2005es}
\bibitem{Riemann:2005es} 
  S.~Riemann,
  %``Z' signals from Kaluza-Klein dark matter,''
  eConf C {\bf 050318}, 0303 (2005)
  [hep-ph/0508136].
  %%CITATION = HEP-PH/0508136;%%
 
%\cite{Hewett:1988xc}
\bibitem{Hewett:1988xc}
  J.~L.~Hewett, T.~G.~Rizzo,
  %``Low-Energy Phenomenology of Superstring Inspired E(6) Models,''
  Phys.\ Rept.\  {\bf 183}, 193 (1989).

%\cite{Erler:2009jh}
\bibitem{Erler:2009jh}
  J.~Erler, P.~Langacker, S.~Munir, E.~Rojas,
  %``Improved Constraints on Z-prime Bosons from Electroweak Precision Data,''
  JHEP {\bf 0908}, 017 (2009)
  [arXiv:0906.2435 [hep-ph]].

%\cite{Langacker:2009su}
\bibitem{Langacker:2009su}
  P.~Langacker,
  %``Z' Physics at the LHC,''
    [arXiv:0911.4294 [hep-ph]].

%\cite{Chatrchyan:2012it}
\bibitem{Chatrchyan:2012it}
  S.~Chatrchyan {\it et al.}  [CMS Collaboration],
%``Search for narrow resonances in dilepton 
%mass spectra in pp collisions at 
%sqrt(s) = 7 TeV,''  
Phys.\ Lett.\ {\bf B714}, 158-179 (2012) 
[arXiv:1206.1849 [hep-ex]]. 
 
\bibitem{atlas-dilepton}
ATLAS Collaboration, Note ATLAS-CONF-2012-007 
(March 2012).
    
%\cite{Osland:2009tn}
\bibitem{Osland:2009tn}
  P.~Osland, A.~A.~Pankov, A.~V.~Tsytrinov, N.~Paver,
  %``Spin and model identification of Z' bosons at the LHC,''
  Phys.\ Rev.\  {\bf D79}, 115021 (2009)
  [arXiv:0904.4857 [hep-ph]].

%\cite{:2007sg}
\bibitem{:2007sg}
 J.~Brau {\it et al.}  [ILC Collaboration],
 ``ILC Reference Design Report Volume 1 - Executive Summary,''
 arXiv:0712.1950 [physics.acc-ph].
 %%CITATION = ARXIV:0712.1950;%%

%\cite{Djouadi:2007ik}
\bibitem{Djouadi:2007ik}
 G.~Aarons {\it et al.}  [ILC Collaboration],
 ``International Linear Collider Reference Design Report Volume 2: PHYSICS AT
 THE ILC,''
 arXiv:0709.1893 [hep-ph].
 %%CITATION = ARXIV:0709.1893;%%

%\cite{Pankov:1990hq}
\bibitem{Pankov:1990hq}
  A.~A.~Pankov, N.~Paver,
  %``Manifestations of heavy extra neutral E(6) gauge bosons in e+ e- ---> W+ W- at LEP-2,''
  Phys.\ Lett.\  {\bf B272}, 425-430 (1991).

%\cite{Pankov:1992cy}
\bibitem{Pankov:1992cy}
  A.~A.~Pankov, N.~Paver,
  %``Probing Z - Z-prime mixing at future e+ e- colliders,''
  Phys.\ Rev.\  {\bf D48}, 63-77 (1993).

%\cite{Pankov:1994hx}
\bibitem{Pankov:1994hx}
  A.~A.~Pankov, N.~Paver,
  %``Initial longitudinal polarization in e+ e- ---> W+ W- as a tool to probe trilinear gauge boson couplings,''
  Phys.\ Lett.\  {\bf B324}, 224-230 (1994).

%\cite{Pankov:1997da}
\bibitem{Pankov:1997da}
  A.~A.~Pankov, N.~Paver, C.~Verzegnassi,
  %``Z-prime effects and anomalous gauge couplings at LC with polarization,''
  Int.\ J.\ Mod.\ Phys.\  {\bf A13}, 1629-1650 (1998)
  [hep-ph/9701359].

%\cite{Jung:1999wq}
\bibitem{Jung:1999wq}
  D.~-W.~Jung, K.~Y.~Lee, H.~S.~Song, C.~Yu,
  %``Polarization effects on $W$ boson pair productions with the extra neutral gauge boson at the $e^{+} e^{-}$ linear collider,''
  J.\ Korean Phys.\ Soc.\  {\bf 36}, 258-264 (2000)
  [hep-ph/9905353].

%\cite{Ananthanarayan:2010bt}
\bibitem{Ananthanarayan:2010bt}
  B.~Ananthanarayan, M.~Patra, P.~Poulose,
  %``Signals of additional Z boson in e+e-\to W+W^- at the ILC with polarized beams,''
  JHEP {\bf 1102}, 043 (2011)
  [arXiv:1012.3566 [hep-ph]].

\bibitem{gounaris}
G. Gounaris, J. L. Kneur, J. Layssac,
G. Moultaka, F. M. Renard and D. Schildknecht, Proceedings of the
Workshop $e^+e^-${\it Collisions at 500 GeV: the Physics
Potential}, Ed. P.M. Zerwas (1992), DESY 92-123B, p.735.

%\cite{Gounaris:1992kp}
\bibitem{Gounaris:1992kp}
  G.~Gounaris, J.~Layssac, G.~Moultaka, F.~M.~Renard,
  %``Analytic expressions of cross-sections, asymmetries and W density matrices for e+ e- ---> W+ W- with general three boson couplings,''
  Int.\ J.\ Mod.\ Phys.\  {\bf A8}, 3285-3320 (1993).

%\cite{MoortgatPick:2005cw}
\bibitem{MoortgatPick:2005cw}
  G.~Moortgat-Pick, T.~Abe, G.~Alexander, B.~Ananthanarayan, A.~A.~Babich, V.~Bharadwaj, D.~Barber, A.~Bartl {\it et al.},
  %``The Role of polarized positrons and electrons in revealing fundamental
  %interactions at the linear collider,''
  Phys.\ Rept.\  {\bf 460}, 131-243 (2008)
  [hep-ph/0507011].

%\cite{Osland:2009dp}
\bibitem{Osland:2009dp}
  P.~Osland, A.~A.~Pankov, A.~V.~Tsytrinov,
  %``Identification of extra neutral gauge bosons at the International Linear Collider,''
  Eur.\ Phys.\ J.\  {\bf C67}, 191-204 (2010)
  [arXiv:0912.2806 [hep-ph]].
    
%\cite{UPR-0476T}
\bibitem{UPR-0476T}
  P.~Langacker and M.~-x.~Luo,
  %``Constraints on additional $Z$ bosons,''
  Phys.\ Rev.\ D\ {\bf 45}, 278  (1992).
  %%CITATION = PHRVA,D45,278;%%

%\cite{Hagiwara:1986vm}
\bibitem{Hagiwara:1986vm} 
  K.~Hagiwara, R.~D.~Peccei, D.~Zeppenfeld and K.~Hikasa,
  %``Probing the Weak Boson Sector in e+ e- ---> W+ W-,''
  Nucl.\ Phys.\ B {\bf 282}, 253 (1987)
  %%CITATION = NUPHA,B282,253;%%

%\cite{Bilenky:1993ms}
\bibitem{Bilenky:1993ms} 
  M.~S.~Bilenky, J.~L.~Kneur, F.~M.~Renard and D.~Schildknecht,
  %``Trilinear couplings among the electroweak vector bosons and their determination at LEP-200,''
  Nucl.\ Phys.\ B {\bf 409}, 22 (1993).
  %%CITATION = NUPHA,B409,22;%%
  
  %\cite{Bilenky:1993uy}
\bibitem{Bilenky:1993uy} 
  M.~S.~Bilenky, J.~L.~Kneur, F.~M.~Renard and D.~Schildknecht,
  %``The Potential of a new linear collider for the measurement of the trilinear couplings among the electroweak vector bosons,''
  Nucl.\ Phys.\ B {\bf 419}, 240 (1994)
  [hep-ph/9312202].
  %%CITATION = HEP-PH/9312202;%%

  %\cite{Nakamura:2010zzi}
\bibitem{Nakamura:2010zzi} 
  K.~Nakamura {\it et al.}  [Particle Data Group Collaboration],
  %``Review of particle physics,''
  J.\ Phys.\ G {\bf 37}, 075021 (2010).
  %%CITATION = JPHGB,G37,075021;%%
  
  %\cite{Fleischer:1991nw}
\bibitem{Fleischer:1991nw} 
  J.~Fleischer, K.~Kolodziej and F.~Jegerlehner,
  %``W pair production in e+ e- annihilation: Radiative corrections including hard bremsstrahlung,''
  Phys.\ Rev.\ D {\bf 47}, 830 (1993).
  %%CITATION = PHRVA,D47,830;%%
  
  %\cite{Beenakker:1994vn}
\bibitem{Beenakker:1994vn} 
  W.~Beenakker and A.~Denner,
  %``Standard model predictions for $W$ pair production in electron - positron collisions,''
  Int.\ J.\ Mod.\ Phys.\ A {\bf 9}, 4837 (1994).
  %%CITATION = IMPAE,A9,4837;%%

%\cite{Denner:2005es}
\bibitem{Denner:2005es} 
  A.~Denner, S.~Dittmaier, M.~Roth and L.~H.~Wieders,
  %``Complete electroweak O(alpha) corrections to charged-current e+e- ---> 4 fermion processes,''
  Phys.\ Lett.\ B {\bf 612}, 223 (2005)
  [Erratum-ibid.\ B {\bf 704}, 667 (2011)]
  [hep-ph/0502063].
  %%CITATION = HEP-PH/0502063;%%
  
%\cite{Denner:2005fg}
\bibitem{Denner:2005fg} 
  A.~Denner, S.~Dittmaier, M.~Roth and L.~H.~Wieders,
  %``Electroweak corrections to charged-current e+ e- ---> 4 fermion processes: Technical details and further results,''
  Nucl.\ Phys.\ B {\bf 724}, 247 (2005)
  [Erratum-ibid.\ B {\bf 854}, 504 (2012)]
  [hep-ph/0505042].
  %%CITATION = HEP-PH/0505042;%%
  
%\cite{Beenakker:1991jk}
\bibitem{Beenakker:1991jk} 
  W.~Beenakker, F.~A.~Berends and T.~Sack,
  %``The Radiative process e+ e- ---> W+ W- gamma,''
  Nucl.\ Phys.\ B {\bf 367}, 287 (1991).
  %%CITATION = NUPHA,B367,287;%%

%\cite{Beenakker:1990sf}
\bibitem{Beenakker:1990sf} 
  W.~Beenakker, K.~Kolodziej and T.~Sack,
  %``The Total cross-section e+ e- ---> W+ W-,''
  Phys.\ Lett.\ B {\bf 258}, 469 (1991).
  %%CITATION = PHLTA,B258,469;%%
  
%\cite{Pankov:2005kd}
\bibitem{Pankov:2005kd}
  A.~A.~Pankov, N.~Paver and A.~V.~Tsytrinov,
  %``Distinguishing new physics scenarios at a linear collider with  polarized
  %beams,''
  Phys.\ Rev.\  D {\bf 73}, 115005 (2006)
  [arXiv:hep-ph/0512131].

\bibitem{LEP2Wpolar}
G. Abbiendi {\it et al.}, [OPAL collaboration], Phys.
Lett. {\bf B585}, 223  (2004);\\
P. Achard {\it et al.}, [L3 collaboration], Phys.
Lett. {\bf B557}, 147  (2003);\\
  J.~Abdallah {\it et al.},  [DELPHI Collaboration],
  %``Study of W boson polarisations and Triple Gauge boson Couplings in the
  %reaction e+e- -> W+W- at LEP 2,''
  Eur.\ Phys.\ J.\  C {\bf 54}, 345 (2008)
  [arXiv:0801.1235 [hep-ex]];\\
J.P. Couchman, A measurement of the triple gauge boson couplings
and $W$ boson polarisation in $W$-pair production at LEP2, Ph.D.
thesis, University College London, 2000.

\end{thebibliography}
\end{document}